\documentclass[a4paper, reqno]{amsart}

\usepackage{latexsym}
\usepackage{amsmath}
\usepackage{amssymb}
\usepackage{amsfonts}
\usepackage{amsthm}
\usepackage{amstext}
\usepackage{enumerate}

\usepackage{pdflscape}

\newcommand{\cQ}{{\mathcal Q}}
      \newcommand{\sgn}{{\operatorname{sgn}}}
      \newcommand{\sah}{{\operatorname{\varsigma}}}
      
      \newcommand{\crit}{{\operatorname{cr}}}
\renewcommand{\bar}{\overline}
      
      \renewcommand{\i}{{\operatorname{i}}}
     \newcommand{\e}{\operatorname{e}}

     \newcommand{\s}{{\operatorname{s}}}
     \renewcommand{\a}{{\operatorname{a}}}
     \renewcommand{\d}{{\operatorname{d}}}
          
     \newcommand{\cl}{{\operatorname{cl}}}
     
     \newcommand{\Ran}{{\operatorname{Ran}}}

\renewcommand{\sp}{{\operatorname{sp}}}

\newcommand{\zz}{{\mathbb{Z}}}
\newcommand{\cc}{{\mathbb{C}}}
\newcommand{\nn}{{\mathbb{N}}}
\newcommand{\rr}{{\mathbb{R}}}
\newcommand{\0}{\mathbf{0}}
\newcommand{\x}{\mathbf{x}}

\newcommand{\y}{\mathbf{y}}
\newcommand{\kk}{{\bf k}}
\newcommand{\q}{{\bf q}}

\newcommand{\cH}{{\mathcal H}}
\newcommand{\cZ}{{\mathcal Z}}

\newcommand{\ess}{{\rm ess}}
\newcommand{\fr}{{\rm fr}}
\newcommand{\exc}{{\epsilon}}
\newcommand{\Exc}{{\rm Exc}}

     \theoremstyle{plain}
     \newtheorem{thm}{Theorem}[section]

     \theoremstyle{definition}

 \newtheorem{conjecture}[thm]{Conjecture}

\def\bbbone{{\mathchoice {\rm 1\mskip-4mu l} {\rm 1\mskip-4mu l}
{\rm 1\mskip-4.5mu l} {\rm 1\mskip-5mu l}}}
\def\one{\bbbone}

\renewcommand{\t}{{\scriptstyle \#}}
\newcommand{\loplus}{\mathop{\oplus}\limits}
\newcommand{\lotimes}{\mathop{\otimes}\limits}
     \numberwithin{equation}{section}
\newpage
\begin{document}
\title{On the 
energy-momentum spectrum of a homogeneous Fermi gas}

\author{Jan Derezi\'{n}ski}
\address[J. Derezi\'{n}ski]
{Dept. of Math. Methods in Phys., Faculty of Physics, University of Warsaw\\ 
Ho{\.z}a 74, 00-682 Warszawa, Poland} 
\email{Jan.Derezinski@fuw.edu.pl}

\author{Krzysztof A. Meissner}
\address[K. A. Meissner]{Institute of Theoretical Physics, Faculty of Physics, University of Warsaw\\
Ho{\.z}a 69, 00-681 Warszawa, Poland}
\email{Krzysztof.Meissner@fuw.edu.pl}

\author{Marcin Napi\'{o}rkowski}
\address[M. Napi\'{o}rkowski]
{Dept. of Math. Methods in Phys., Faculty of Physics, University of Warsaw\\ 
Ho{\.z}a 74, 00-682 Warszawa, Poland} 
\email{Marcin.Napiorkowski@fuw.edu.pl}

\date{\today}

\begin{abstract} 
We consider 
 translation invariant quantum systems in thermodynamic limit.
We argue that their energy-momentum spectra should have   shapes consistent with effective models involving quasiparticles.
Our main example is
 second quantized homogeneous interacting 
Fermi gas in a large cubic box with periodic boundary conditions, 
at zero temperature.  We expect that its energy-momentum spectrum has  a positive energy gap and a positive critical velocity.

\end{abstract}
\maketitle
\tableofcontents

\section{Introduction}
\label{s1}

\subsection{Excitation spectrum of Fermi gas}

In \cite{CDZ}, one of the authors of this paper together with H.~Cornean and
P.~Zi\'{n}{} discussed a number of conjectures about
the {\em excitation spectrum}  of the  interacting Bose gas at
zero temperature with repulsive
potentials in the thermodynamic limit. 
In particular,  \cite{CDZ} conjectured that such systems have a quasiparticle-like excitation spectrum without an energy gap and with a positive critical velocity. 
These conjectures seem to be
consistent with experimental data. In particular, they explain 
various phenomena related to the {\em superfluidity}.

In this paper we would like to sketch a number of
 analogous conjectures about
the interacting
 Fermi gas at zero temperature with an attractive interaction in thermdynamic limit. We will argue that these systems should possess a quasiparticle-like excitation spectrum with a positive energy gap and a positive critical velocity. 
This conjecture implies in particular that the ground
state energy is separated from the rest of the spectrum. In some
situations, this ground state can be interpreted as a {\em current
carrying state}, and plays an important role in
 the phenomenon of {\em superconductivity}.

The most robust quantity related to the excitation spectrum seems to be its
infimum. Therefore, our main conjectures involve the infimum of the excitation
spectrum (separately in the even and odd sector). They are based on the
{\em grand-canonical Hamiltonian} for a fixed {\em chemical potential}.

However, 
to have full information about the excitation
spectrum, it is not enough to know its infimum. In fact, the HFB approximation
suggests that the excitation spectrum of the Fermi gas  has
``lacunas'' near its bottom.
 We make an attempt to express some conjectures about  these lacunas.
 These conjectures are more
complicated to state, probably also more delicate, and involve the {\em canonical
Hamiltonian},  (that is, for a fixed number of particles).


We do not prove our conjectures. However, we introduce a rather general class
of model Hamiltonians for which these conjectures can be tested and, maybe,   proven under some assumptions.
These Hamiltonians consist of a two-body kinetic energy, not necessarily quadratic, and an interaction. 
The interaction does not have to be local (given by a local potential).
We also suppose that our system has ``internal degrees of
freedom'' (eg. ``spin'').
It seems important to assume that the interaction is in some sense
``attractive'', which means some kind of negative definiteness. This is suggested by the Hartree-Fock-Bogoliubov (HFB) method.

We include in our paper short computations based on  the HFB approach. The
basic principle of these computations is well known, but 
 in the literature they are usually presented
in simple special cases. Our presentation applies to a rather general case. As a result of the HFB approach we obtain an approximate quasiparticle representation of our Hamiltonian with a dispersion relation possessing very special features: a positive energy gap and a positive critical velocity.
We  conjecture  that this dispersion relation  suggests basic qualitative features of the true excitation spectrum of interacting Fermi gas in thermodynamic limit. 

\subsection{Role of translation invariance}

There exist many papers that study the energy spectrum of interacting Fermi and Bose
systems. In particular, there exist interesting works that study the 
HBF approximation in such systems. What makes our paper different is the role
of  translation invariance. This enables us to ask questions about the
excitation spectrum,  which we expect to have interesting properties.

Many papers attempt to show that models based on quasiparticles give some kind of an approximation to realistic Hamiltonians, see eg. \cite{CS,BBZKT}. However,  only relatively crude features are  considered
in essentially all these papers. Typically, they study the energy or the free energy per volume in thermodynamic limit. We are interested in the excitation spectrum, which is a finer quantity and does not involve dividing by the large volume. The only rigorous result that we know devoted to the excitation spectrum of an interacting quantum gas is due to R.~Seiringer \cite{Sei}. It concerns the Bose gas in finite volume and a mean field limit.

We always assume that the interaction is
translation invariant. 
In realistic physical systems translation invariance is at most approximate.
 Superconducting materials are perhaps the closest to  idealized
translation invariant models that we consider.
Nevertheless, we believe that the picture presented in our paper is physically
relevant also in many Fermi systems that are quite far from being translationally
invariant, 
 such as
 quantum dots and nuclei. In fact,  one can argue that ``traces'' of
 translational degrees of freedom are present also in these systems,
disguised as rotations and vibrations. In particular,
the energy spectra of their odd and even sectors have various features consistent
with our conjectures, see
\cite{BRT} for quantum dots and \cite{RS} Sect. 6.1 for nuclei.

The immediate motivation of our paper is to state
mathematically interesting  rigorous conjectures together with heuristic
arguments in their favor. Therefore, we do not strive at all costs
 to describe
 realistic concrete physical systems. The assumption of translation invariance helps to formulate a clean and
 rigorous 
 definition of various concepts.

\noindent{\bf Acknowledgement}
The research of J.D. and M.N was supported in part by the National
Science
 Center (NCN) grant No.  2011/01/B/ST1/04929.
The work of M.N. was also supported by the Foundation for Polish
Science International PhD Projects Programme co-financed by the EU
European Regional Development Fund.

\section{Quasiparticles and quasiparticle-like excitation spectrum}

The concept  of ``quasiparticle'',
although often used,  seems to have no satisfactory definition  in the literature. In this section we  attempt to give a number of rigorous interpretations of this term. We will also discuss spectral properties of  quantum systems that can be described in terms of quasiparticles.

The discussion of this section will be rather general and abstract. To a large extent it will be independent of the rest of the paper.

\subsection{Translation invariant quantum systems}

The main object of interest of this paper are translation invariant quantum systems in {\em thermodynamic limit}. There are at least two approaches that can be used to describe such systems.

In the first approach one starts with a construction of  a
 system in finite volume, using 
$\Lambda=[-L/2,L/2]^d$, the $d$-dimensional
 cubic box of side length $L$, as the configuration space.
It is convenient, although somewhat unphysical, to impose the periodic boundary conditions,  The system is described by its Hilbert space $\cH^L$,
 Hamiltonian $H^L$  and momentum $P^L$. The spectrum of the momentum is discrete and coincides with $\frac{2\pi}{L}\zz^d$. 
After computing appropriate quantities (such as the infimum of the excitation spectrum) one tries to take the limit $L\to\infty$.

Sometimes a different approach is possible. One can try to construct a Hilbert space $\cH$, a Hamiltonian $H$ and a momentum $P$ that describe the system  on $\rr^d$. This may be not easy. It may require the use of refined techniques \cite{BR2,GJ}. It is probably not always  possible.
Note that in this case the spectrum of the momentum is   expected to be absolutely continuous, with the exception of the ground state.

The latter approach seems conceptually more elegant. Throughout most of this section we will adopt it. In most situations this will allow us to formulate some of the physical concepts in a concise manner. (Sometimes, however, it will lead to technical complications).

In the next two sections we adopt the former approach, which is more down-to-earth. Thus only a family $(H^L,P^L)$ for finite $L$ will be defined.

To sum up,
throughout  most of this section 
by a {\em translation invariant quantum system} we  will mean
$d+1$ commuting self-adjoint operators $(H,P_1,\dots,P_d)$ on a Hilbert space $\cH$. $H$ has the interpretation of a {\em Hamiltonian} and $P=(P_1,\dots, P_d)$ describes the {\em momentum}.

\subsection{Excitation spectrum}
\label{Excitation spectrum}

The joint  spectrum of the operators $(H,P)$  (which is a subset of $\rr^{1+d}$) will be denoted by $\sp(H,P)$ and called the  {\em energy-momentum spectrum of $(H,P)$}.

We will often assume that $H$ is bounded from below. If it is the case, we can define the {\em ground state energy} as $E:=\inf\sp H$. 
We will also often assume that
$H$ possesses translation invariant ground state $\Phi$, which is a unique joint eigenvector of $H,P$. In particular, $H\Phi=E\Phi$ and $P\Phi=0$.

Under these assumptions, by subtracting the ground state energy from the  energy-momentum spectrum we obtain the {\em excitation spectrum of $(H,P)$}, that is, $\sp(H-E,P)$.
We can also introduce the {\em strict excitation spectrum} as the joint spectrum of restriction of $(H-E,P)$ to the orthogonal complement of $\Phi$:
\begin{equation}\Exc:=\sp\Bigl((H-E,P)\Big|_{\{\Phi\}^\perp}\Bigr)\label{exci-1}
\end{equation}
Thus if $(E,\0)$ is an isolated simple eigenvalue of $(H,P)$, then 
\[\Exc=\sp(H-E,P)\backslash(0,\0).\]
Otherwise $\Exc=\sp(H-E,P)$.

We introduce also a special notation for the infimum of $\Exc$:
\begin{eqnarray*}
\exc(\kk)&:=&\inf\{e\ :\ (e,\kk)\in\Exc\}.
\end{eqnarray*}

The following two parameters 
 have interesting
physical implications. 
The first is  the {\em energy gap}, defined as
\[\varepsilon:=\inf\Bigl(\sp (H-E)\Big|_{\{\Phi\}^\perp}\Bigr)=\inf\{\exc(\kk)\ :\ \kk\in\rr^d\}.\]
Another quantity of physical interest is the {\em critical velocity}:
\[c_\crit:=\inf_{\kk\neq\0}\frac{\epsilon(\kk)}{|\kk|}.\]

Physical properties of a system are especially interesting if the energy gap $\varepsilon$
is strictly positive. In such a case, the ground state energy
 is separated from the
rest of the energy spectrum, and hence the ground state is {\em stable}.

Positive critical velocity is also very interesting. 
Physically,  a positive critical velocity is closely related to the phenomenon
of 
{\em superfluidity}, see eg. a discussion in  \cite{CDZ}.

\subsection{Essential excitation spectrum}
\label{Essential excitation spectrum}

One expects that most of a typical  excitation spectrum is absolutely continuous wrt. the Lebesgue measure on $\rr^{d+1}$. However, it  may also contain  isolated shells 
 continuously depending on the momentum. In this subsection we attempt to define  the part of the excitation spectrum that corresponds to such a situation. 

Note that this is not easy in the abstract framework that we adopted in this section. Actually, in the next section, based on finite volume systems $(H^L,P^L)$, we will use a different approach to  define isolated shells, see Subsect \ref{Excited quasiparticle shells}.

We say that $(e,\kk)\in \rr^{d+1}$ belongs to $\Exc_\d$, called the
 {\em discrete excitation spectrum}, if there exists
 $\delta>0$ such that the operator $P$ has an absolutely continuous spectrum of uniformly  finite multiplicity when restricted to \[\Ran\,\one(|H-E-e|<\delta)\one(|P-\kk|<\delta).\]
The {\em essential excitation spectrum} is defined as $\Exc_\ess:=\Exc\backslash \Exc_\d$.

(We use an obvious notation for spectral projections of self-adjoint operators $H$ and $P$: eg. $\one(|H-e|<\delta)$ denotes the spectral projection of $H$ onto $]e-\delta,e+\delta[$).

We introduce also a special notation for the bottom of  $\Exc_\ess$:
\begin{eqnarray*}
\exc_\ess(\kk)&:=&\inf\{e\ :\ (e,\kk)\in\Exc_\ess\}.
\end{eqnarray*}

Obviously,
\[\Exc\supset
\Exc_\ess,\]
\[\exc(\kk)\leq\exc_\ess(\kk),\ \ \kk\in\rr^d.\]

Note that   typically   $\Exc_\d$ consists of 
a finite number of  shells separated by
lacunas.

Abstract theory allows us to
 represent the Hilbert space $\cH$ as the direct integral over $\rr^d$ given by the spectral decomposition of $P$, see eg. \cite{BR1}, 4.4.1. Suppose, in addition, that  this direct integral can be taken with respect to the Lebesgue measure, so that we can write
\begin{equation}H=\int\limits_{\kk\in\rr^d}^\oplus H(\kk)\d\kk.\label{pw0}\end{equation}
Then  it is tempting to claim that
\begin{eqnarray}
\sp(H-E,P)&=&\bigcup\limits_{\kk\in\rr^d}\sp\bigl( H(\kk)-E\bigr)\times\{\kk\},\label{pw1}\\
\Exc_\ess&=&\Bigl(\bigcup\limits_{\kk\in\rr^d}\sp_\ess \bigl(H(\kk)-E\bigr)\times\{\kk\}\Bigr)^\cl,\label{pw2}
\end{eqnarray}
where $\sp_\ess$ denotes the essential spectrum and the subscript $\cl$ denots the closure.
Unfortunately, at this level of generality there is a problem with (\ref{pw1}) and (\ref{pw2}). First of all, there is no guarantee that we can put the Lebesgue measure in (\ref{pw0}). Secondly, the direct integral representation (\ref{pw0}) is not defined uniquely, but only modulo sets of measure zero.

 In concrete situations, however,
 (such as quasiparticle systems considered in Sect.
\ref{Concept of a quasiparticle})
the direct integral (\ref{pw0}) has an obvious distinguished realization involving the Lebesgue measure, for which the identities (\ref{pw1}) and (\ref{pw2}) are actually true.

\subsection{Quasiparticle quantum systems}
\label{Concept of a quasiparticle}

Many important  translation invariant
 quantum systems can be described in terms of 
 {\em quasiparticles}, that is,  independent
 bosonic or fermionic  modes with appropriately
chosen {\em dispersion relations}  (the dependence of the
 quasiparticle energy on the momentum).

Let us be more precise.  For a Hilbert space $\cZ$, the notation $\Gamma_\s(\cZ)$, resp. $\Gamma_\a(\cZ)$ will stand for the {\em bosonic}, resp. {\em fermionic Fock space} with the {\em one particle space} $\cZ$. 

By a  {\em quasiparticle quantum system} we will mean
$(H_\fr,P_\fr)$, where
\begin{eqnarray}
H_\fr&=&\sum_{i\in\cQ}\int_{I_i}\omega_i(\kk)b_i^*(\kk)b_i(\kk)\d\kk ,
\label{quasa1} \\
P_\fr&=&\sum_{i\in\cQ}\int_{I_i}\kk b_i^*(\kk)b_i(\kk)\d\kk ,
\label{quasa2}
\end{eqnarray}
for some intervals $I_i\subset\rr^d$, real continuous functions $I_i\ni\kk\mapsto\omega_i(\kk)$, and {\em creation}, resp. {\em annihilation operators}  $b_i^*(\kk)$ and $b_i(\kk)$.
 $\cQ$ is called the set of  {\em quasiparticle species} and it is partitioned into $\cQ_\s$ and $\cQ_\a$ -- bosonic and fermionic quasiparticles. 

 We are using the standard notation of the formalism of 2nd quantization:  $b_i^*(\kk)$ and $b_i(\kk)$ satisfy the usual commutation/anticommutation relations. They  are not true operators, only formal symbols, however the right hand sides of
(\ref{quasa1}) and (\ref{quasa2}) are  well defined as operators on the Fock space
\begin{equation}
\lotimes_{i\in\cQ_\s}\Gamma_\s\left(L^2(I_i)\right)\otimes
\lotimes_{j\in\cQ_\a}\Gamma_\a\left(L^2(I_j)\right).\label{fock}\end{equation}

 For $i\in\cQ$,
the set $I_i$ describes the allowed  range of the momentum of a single $i$th quasiparticle and 
 $\omega_i(\kk)$ is its energy  (dispersion relation)
for momentum $\kk\in\rr^d$. Note that $I_i$ can be strictly smaller than $\rr^d$ -- some quasiparticles may exist only for some momenta.
 This allows us more flexibility and is consistent with  applications to condensed matter physics. 
 It will be convenient to define \[I(\kk):=\{i\in\cQ\ :\ \kk\in I_i(\kk)\}\] (the set of quasiparticles that may have momentum  $\kk\in\rr^d$).

Clearly, if we know the dispersion relations
 $I_i\ni\kk\mapsto \omega_i(\kk)$, $i\in\cQ$, then we can determine 
the energy-momentum spectrum of $(H_\fr,P_\fr)$:
\begin{eqnarray*}
\sp
(H_\fr,P_\fr)&=&\{(0,\0)\}\\
&&\!\!\cup\ \bigl\{\bigl(\omega_{i_1}(\kk_1)+\cdots+\omega_{i_n}(\kk_n),\kk_1+\cdots+\kk_n\bigr)\ :
\ n=1,2,3\dots\}^\cl
.\end{eqnarray*}
Note that there is an obvious direct integral representation of the form (\ref{pw0}) and the relations (\ref{pw1}) and (\ref{pw2}) hold.

\subsection{Properties of the excitation spectrum of quasiparticle systems}

Let $(H,P)$ be a  quasiparticle  system.
The energy-momentum spectrum of such systems has special properties. 
First, we have 
\begin{eqnarray}
(0,\0)&\in&\sp(H,P),\label{endeq0-}\end{eqnarray} 
because of the Fock vacuum state, which is a unique joint eigenstate of $(H,P)$.  Moreover, we have a remarkable addition property
\begin{eqnarray}
\sp(H,P)&=&\sp(H,P)+\sp(H,P).\label{endeq1-}\end{eqnarray} 

Assume now that
the Hamiltonian (\ref{quasa1}) is bounded from below, or what is equivalent,  assume that all the dispersion relations are non-negative.
 Then the Fock vacuum is a ground state satisfying $E=0$, so that  the excitation spectrum coincides with the energy-momentum spectrum.
Thus
we can rewrite (\ref{endeq0-}) and
(\ref{endeq1-}) as
\begin{eqnarray}
(0,\0)&\in&\sp(H-E,P),\label{endeq0}\\
\sp(H-E,P)&=&\sp(H-E,P)+\sp(H-E,P).\label{endeq1}\end{eqnarray}

Given (\ref{endeq0}), (\ref{endeq1}) is equivalent to
\begin{eqnarray}
\Exc&\supset&\Exc+\Exc.\label{endeq1.}\end{eqnarray}

Another remarkable property holds true if in addition the number of particle species is finite. We have  then
\begin{eqnarray}
\Exc_\ess&=&\bigl(\Exc+\Exc\bigr)^\cl.\label{endeq4}\end{eqnarray}
Indeed, using the continuity of the momentum spectrum, we easily see that only 1-particle states can belong to the disrete spectrum of the fiber Hamiltonians $H(\kk)$.

Before we proceed, let us introduce some terminology
concerning  real functions that will be useful in our study of quasiparticle-like spectra.
Recall that a function $\rr^d\ni\kk\mapsto\epsilon(\kk)$ is called {\em
  subadditive} 
if
\[\epsilon(\kk_1+\kk_2)\leq
\epsilon(\kk_1)+\epsilon(\kk_2).\]

Let 
$\rr^d\supset I\ni\kk\mapsto\omega(\kk)$ be a given function.
Define
\begin{eqnarray}
\sah_\omega(\kk)&=&\inf
\{\omega(\kk_1)+\cdots+\omega(\kk_n)\ :\ \kk_1+\cdots+\kk_n=\kk,
\ n=1,2,3,\dots\}, \nonumber
\\
\sah_{\ess,\omega}(\kk)&=&\inf
\{\omega(\kk_1)+\cdots+\omega(\kk_n)\ :\ \kk_1+\cdots+\kk_n=\kk,
\ n=2,3,\dots\}, \nonumber
\end{eqnarray}
(By definition, the infimum of an empty set is $+\infty$).
 $\sah_\omega$ is known under the name of the {\em 
subadditive hull of $\omega$}. Equivalently, 
 $\sah_\omega$ is the biggest subadditive function less than $\omega$.

Note the relation
\begin{eqnarray*}
\sah_\omega(\kk)&=&\min\{\omega(\kk),\sah_{\ess,\omega}(\kk)\}.
\end{eqnarray*}

Let us go back to a quasiparticle  system (\ref{quasa1}), (\ref{quasa2}) with nonnegative dispersion relations.
For $\kk\in\rr^d$, define
\begin{equation}
  \omega_{\min}(\kk):=\min\{\omega_i\ :\ i\in I(\kk)\}.\label{mino}\end{equation}
Recall the functions  $\exc$ and $\exc_\ess$  and the parameters
$\varepsilon$ and $c_\crit$ that we defined in Subsects 
\ref{Excitation spectrum} and \ref{Essential excitation spectrum}.

\begin{thm} 
\begin{enumerate}
\item The bottom of the strict excitation spectrum is the subadditive hull of $\omega_{\min}$:
\begin{eqnarray}
\exc(\kk)&=&
\sah_{\omega_{\min}}(\kk),\ \kk\in\rr^d.\nonumber
\end{eqnarray}
\item The energy gap satisfies
\begin{eqnarray*}
\varepsilon&=&\inf_\kk\omega_{\min}(\kk).
\end{eqnarray*}
\item The critical velocity satisfies
\begin{eqnarray*}
c_\crit&=&\inf_{\kk\neq\0}\frac{\omega_{\min}(\kk)}{|\kk|}
=\inf_{\kk\neq\0}\frac{\epsilon_\ess(\kk)}{|\kk|}.
\end{eqnarray*}
\item 
If in addition the number of quasiparticle species is finite, then
\begin{eqnarray} \exc_\ess(\kk)&=&\sah_{\ess,\omega_{\min}}(\kk),\ \ \kk\in\rr^d.\nonumber
\end{eqnarray}
\end{enumerate}
\end{thm}

Note that we assume that the momentum space is $\rr^d$. If we replace the momentum space
$\rr^d$ with $\frac{2\pi}{L}\zz^d$ (that is, if we put 
our system on a torus of side length $L$)  and we assume that all quasiparticles are bosonic, then
all statements of this
subsection generalize in an obvious way. However, because of the Pauli
principle, not all of them generalize  in
the  fermionic case.

\subsection{Approximate versus exact quasiparticles}
\label{Approximate versus exact quasiparticles}

  One often considers quantum systems of the form
\begin{equation}
H=H_\fr +V,\ \ \ P=P_\fr,
\end{equation}
where $(H_\fr,P_\fr)$ is a quasiparticle  system and  the perturbation $V$  is in some sense small. 
A description of physical systems in terms of approximate quasiparticles is very common in condensed matter physics. In particular, it appears naturally in the context of  the so-called Hartree-Fock-Bogoliubov approximation, where one tries to optimize a quasiparticle description for a given quantum system \cite{DNS}.

Clearly, there is a considerable freedom in choosing the splitting of $H$ into $H_\fr$ and $V$, and so   quasiparticles of this kind are only vaguely determined.
We will argue that in some cases   a different concept of quasiparticles is useful, which is rigorous and in a way much more interesting. This concept is expressed in the following definition.

Let $(H,P)$ be a translation invariant  system on a Hilbert space $\cH$.
We will say that it is a {\em   quasiparticle-like system} if it is unitarily equivalent to a quasiparticle  system.

\subsection{Asymptotic quasiparticles}

The above definition has one drawback. In practice we expect that the unitary equivalence mentioned in this definition is in some sense {\em natural} and constructed in the framework of {\em scattering theory}. 

Scattering theory is quite far from the main subject of this paper, which is mostly concerned with purely spectral questions. However, since it has been mentioned and is very closely related to the concept of a quasiparticle, let us give a brief discussion of this topic.

For a number of many body systems the basic idea  of  scattering theory can be described as follows. Using
 the evolution $\e^{\i tH}$ for $t\to\pm\infty$, we define  two  isometric operators
\begin{equation}S^\pm:\lotimes_{i\in\cQ_\s}\Gamma_\s\left(L^2(I_i)\right)\otimes
\lotimes_{j\in\cQ_\a}\Gamma_\a\left(L^2(I_j)\right)\to\cH.
\label{paha}\end{equation}
$S^\pm$ are called the {\em wave} or {\em M{\o}ller operators} and they satisfy
\[HS^\pm=S^\pm H_\fr,\ \ \ PS^\pm=S^\pm P_\fr,\]
where $(H_\fr,P_\fr)$ is  a quasiparticle  system.
 $S:=S^{+*}S^-$ is then called the {\em scattering operator}.

We will say that the system is {\em asymptotically complete} if the wave operators $S^\pm$ are unitary. Clearly, if a system is asymptotically complete, then it is  quasiparticle-like.

 There are at least two 
 classes of important physical system which possess a natural and rigorous scattering theory of this kind.

The first class
 consists of the 2nd quantization of Schr\"odinger many body operators with  2-body  short range interactions  \cite{De2}. One can show that these systems are asymptotically complete
 (see \cite{De1} and references therein).
In this case the system is invariant wrt. the {\em Galileian group}
and the
dispersion relations have the form $\rr^d\ni\kk\mapsto E+\frac{\kk^2}{2m}$. 
Quasiparticles obtained in this context can be ``elementary'' -- in applications to physics these are typically electrons and nuclei -- as well as ``composite'' -- atoms, ions, molecules, etc.

Another important class of systems where the concept of asymptotic quasiparticles has a rigorous foundation belongs to (relativistic) quantum field teory, as axiomatized by the  Haag-Kastler or Wightman axioms. If we assume the existence 
 of   discrete mass shells, 
the so-called {\em Haag-Ruelle theory} allows us to construct  the wave operators, see eg. \cite{Jost}.
Note that in this case the system is covariant wrt. the {\em Poincar\'e{} group} and the dispersion relation has the form $\rr^d\ni \kk\mapsto\sqrt{m^2+\kk^2}$. 
 Here, quasiparticles are the usual stable particles.

Let us stress that both classes of systems  can be interacting in spite of the fact that they are equivalent to  free quasiparticle systems. In particular, their scattering operator can be nontrivial.

The above described  classes of quantum systems 
 are quite special. They are covariant wrt. rather large groups (Galilei or Poincar\'e) and have quite special dispersion relations.

\subsection{Quasiparticles in condensed matter physics}

The concept of a quasiparticle  is useful also in other  contexts, without the Galilei or Poincar\'{e}{} covariance.

An  interesting system which admits a quasiparticle interpretation is the free Fermi gas with a positive chemical potential. We describe this system in Subsect.
\ref{Non-interacting Fermi gas}. In this case the scattering theory is trivial:  $S^+=S^-$, and hence $S=\one$.

It seems that condensed matter physicists apply successfully the concept of a quasiparticle also to various interacting  translation invariant  systems.

One class of such systems seems to be the  Bose gas with repulsive interactions at zero
temperature and positive density. In  this case, apparently, the system is 
typically  well described by a free Bose gas of
 quasiparticles of (at least) two kinds: at low momenta we have {\em phonons} with an approximately linear dispersion relation,
 and at somewhat higher momenta  we have  {\em rotons}.
This idea underlies the famous {\em Bogoliubov approximation} \cite{Bog}, see also
\cite{FW,CDZ}. The phenomenon of
{\em superfluidity} can be to a large extent explained within this picture.
 The model of free asymptotic phonons 
seems to work well in real experiments \cite{M}.

Another class of strongly interacting systems  that seems to be
successfully  modelled by independent quasiparticles
  is the 
Fermi gas  with attractive interactions at zero
temperature and positive chemical potential. 
By using the {\em Hartree-Fock-Bogoliubov (HFB) approach} \cite{RS},
 which is closely related to the original {\em Bardeen-Cooper-Schrieffer (BCS)
approximation} \cite{BCS},
one obtains a simple model that can be used to explain
the {\em superconductivity} of the Fermi gas at very low temperatures. The corresponding quasiparticles are sometimes called {\em partiholes}.

Note that the above two examples -- the interacting Bose and Fermi gas -- are 
 neither Galilei nor Poincare covariant. This  allows us to consider more general dispersion relations. However, we do not know whether these systems admit a quasiparticle interpretation or possess some kind of scattering theory.
Unfortunately, rigorous results in this direction are rather modest.  (There are  attempts at scattering theory for some non-relativistic models of quantum field theory, see \cite{DG1} and \cite{FGS}. There exist also some results
 in a purely perturbative approach \cite{Sch}).

\subsection{Quasiparticle-like excitation spectrum}
\label{Quasiparticle-like excitation spectrum}

The concept of a quasiparticle-like system, as defined in Subsect.
\ref{Approximate versus exact quasiparticles},
is probably too strong for many applications. Let us propose a weaker property, which is  more likely to be satisfied in various situations. 

Again, our starting point is a translation invariant system described by its Hamiltonian and momentum $(H,P)$. Let us assume that $H$ is bounded from below, with $E$, as usual, denoting the ground state energy.
 We will say that the {\em excitation spectrum of  $(H,P)$ is quasiparticle-like} if it coincides with the excitation spectrum
 of a quasiparticle  system (see
 (\ref{quasa1}) and (\ref{quasa2})).
Clearly, the
 excitation spectrum of   a quasiparticle-like system with a bounded from below Hamiltonian  is  quasiparticle-like. However, a system may have a quasiparticle-like excitation spectrum without being a quasiparticle-like system.

A quasiparticle-like excitation spectrum has special properties. In particular, it satisfies  (\ref{endeq0}) and (\ref{endeq1}).

There exists a heuristic, but, we believe, a relatively convincing general argument why realistic translation invariant
quantum systems in thermodynamic limit  at zero temperature should satisfy 
 (\ref{endeq0}) and (\ref{endeq1}).
 It was essentially described at length in \cite{CDZ}, but for the convenience of the reader we reproduce it here. Note in particular that 
it 
the infinite size of the quantum system plays an important role in this argument.

Consider a quantum gas in a  box of a very large side length $L$, described by $(H^L,P^L)$. For shortness, let us drop the superscript $L$. 
First of all, it seems reasonable to assume that the system possesses a translation invariant ground state, which we will denote by $\Phi$, so that $H\Phi=E\Phi$, $P\Phi=0$. Thus (\ref{endeq0}) holds.

Let $(E_{}+e_i,\kk_i)\in\sp(H,P)$, $i=1,2$. We can find eigenvectors with these
eigenvalues, that is, vectors $\Phi_i$ satisfying $H\Phi_i=(E_{}+e_i)\Phi_i$,
$P\Phi_i=\kk_i\Phi_i$.  Let us make the assumption that  it is possible to find
operators $A_i$ that are polynomials in creation and annihilation operator
smeared with functions well localized in  configuration space such that
$PA_i\approx A_i(P+\kk_i)$,  and which approximately 
create the vectors $\Phi_i$ from the
ground state, that is
 $\Phi_i\approx A_i\Phi_{}$.
(Note that here a large size of $L$ plays a role).
 By replacing $\Phi_2$ with $\e^{\i \y P}\Phi_2$ for
some $\y$ and $A_2$ with $\e^{\i\y P}A_2 \e^{-\i\y P}$, we can make sure that the
regions of localization of $A_1$ and $A_2$  are separated by a large distance.

 Now consider the vector $\Phi_{12}:=
A_1A_2\Phi_{}$. Clearly, \[P\Phi_{12}\approx (\kk_1+\kk_2)\Phi_{12}.\]
$\Phi_{12}$ looks like the vector $\Phi_i$ in
 the region of localization of $A_i$, elsewhere it looks like $\Phi_{}$.
 The
Hamiltonian $H$  involves only
 expressions of short range (the potential decays in space). Therefore, 
we expect that
\[H\Phi_{12}\approx (E_{}+e_1+e_2)\Phi_{12}.\]
If this is the case, it
 implies
 that $(E_{}+e_1+e_2,\kk_1+\kk_2)\in\sp(H,P)$.
Thus  (\ref{endeq1.}) holds.

\subsection{Bottom of a quasiparticle-like excitation spectrum}

Now suppose that 
$(H,P)$ is an arbitrary translation invariant  system with a bounded from below Hamiltonian. For simplicity, assume that its ground state energy is zero.
 We assume that we know  its excitation spectrum $\sp(H,P)$.
There are
two natural questions
\begin{enumerate}
\item Is $\sp(H,P)$ quasiparticle-like?
\item If it is the case, to what extent its dispersion relations are determined uniquely?
\end{enumerate}

In order to give partial answers to the above questions, recall 
the functions $\exc$ and $\exc_\ess$, as well as the sets $\Exc_\d$ and $\Exc_\ess$
that we defined in Subsects 
\ref{Excitation spectrum} and \ref{Essential excitation spectrum}.

\begin{thm} Suppose that the excitation spectrum of $(H,P)$ is quasiparticle-like. Then the following is true:
\begin{enumerate}
\item 
 $\epsilon$ is subadditive.
\item
We can partly reconstruct some of the dispersion relations:
\begin{equation}\Exc_\d=\{(\omega_i(\kk),\kk)\ : \ i\in\cQ,\ \kk\in\rr^d\}\,\setminus\, \Exc_\ess
.\label{paio}\end{equation}
Consequently,
for $\kk$ satisfying $\epsilon(\kk)<\epsilon_\ess(\kk)$,
\[\epsilon(\kk)=\omega_{\min}(\kk),\]
where $\omega_{\min}$ was defined in (\ref{mino}).
\item If the number of quasiparticles species is finite, we can reconstruct 
$\epsilon_\ess$  from $\epsilon$:
\begin{eqnarray}
\epsilon_\ess(\kk)&=&\inf \{\epsilon(\kk_1)+
\epsilon(\kk_2)\ :\ \kk=\kk_1+\kk_2\}.\label{essop}
\end{eqnarray}
\end{enumerate}
\end{thm}

The existential part of the inverse problem has a partial solution:
\begin{thm}
Suppose that  $\rr^d\ni\kk\mapsto\omega(\kk)$  be a given subbadditive function. Consider the translation invariant system
\[
 H_\fr=\int\omega(\kk)b_\kk^*b_\kk\d\kk,\ \ \ 
 P_\fr=\int \kk b_\kk^*
b_\kk \d\kk.\]
Then 
\begin{eqnarray*}\epsilon(\kk)&=&\omega(\kk),\\
\epsilon_\ess(\kk)&=&\inf \{\omega(\kk_1)+
\omega(\kk_2)\ :\ \kk=\kk_1+\kk_2\}.
\end{eqnarray*}
\end{thm}

The answer to the uniqueness part of the inverse problem is negative. The only situation where we can identify
dispersion relations from the spectral information involves
$\Exc_\d$, see (\ref{paio}).
The following example shows that we have quite a lot of freedom in choosing a dispersion relation giving a prescribed excitation spectrum.
For instance, all the Hamiltonians below have the same excitation spectrum and essential excitation spectrum  with  $\epsilon(\kk)=\epsilon_\ess(\kk)=|\kk|$:
\[H=\int_{|\kk|<c}|\kk|(1+d|\kk|^\alpha)b_\kk^*b_\kk\d\kk,\]
where $c>0$, $d\geq0$ and $\alpha>0$ are arbitrary.

\subsection{Translation invariant  systems with  two superselection sectors}

Suppose that a Hilbert space $\cH$ has a decomposition $\cH=\cH^+\oplus\cH^-$, which can be treated as a {\em superselection rule}. This means that all observables decompose into direct sums. In particular, the Hamiltonian and momentum decompose as 
$(H,P)=(H^+,P^+)\oplus( H^-,P^-)$.
Clearly, 
\begin{eqnarray}
\sp(H,P)&=&\sp(H^+,P^+)\cup\sp(H^-,P^-).
\end{eqnarray}

 We will often assume that $H$ is bounded from below and possesses a translation invariant ground state $\Phi$ with energy $E$, which belongs to the sector $\cH^+$. The sector $\cH^+$ will be called {\em even}. The other sector $\cH^-$ will be called {\em odd}.

Under these assumptions we will call $\sp(H^+-E,P^+)$, resp.  $\sp(H^--E,P^-)$ the {\em even}, resp. {\em odd excitation spectrum}.
We introduce also the {\em strict even excitation spectrum}:
\begin{equation}\Exc^+:=\sp\Bigl((H^+-E,P^+)\Big|_{\{\Phi\}^\perp}\Bigr)\label{exci-2}
\end{equation}
The  {\em strict odd excitation spectrum} will coincide with
the full odd excitation spectrum:
\begin{equation}\Exc^-:=\sp(H^--E,P^-).\label{exci-3}
\end{equation}

Finally, we define the {\em even} and {\em odd essential  excitation spectrum}
$\Exc_\ess^\pm$ just as in Subsect. 
\ref{Essential excitation spectrum}, except that we replace $(H,P)$ with $(H^\pm,P^\pm)$.

We introduce also a special notation for the bottom of the sets $\Exc^\pm$ and $\Exc_\ess^\pm$:
\begin{eqnarray*}
\exc^\pm(\kk)&:=&\inf\{e\ :\ (e,\kk)\in\Exc^\pm\},\\
\exc_\ess^\pm(\kk)&:=&\inf\{e\ :\ (e,\kk)\in\Exc_\ess^\pm\}.
\end{eqnarray*}

Clearly,
\begin{eqnarray}
\sp(H-E,P)&=&\sp(H^+-E,P^+)\cup\sp(H^--E,P^-),\\
\Exc&=&\Exc^+\cup\Exc^-,\\
\Exc_\ess&=&\Exc_\ess^+\cup\Exc_\ess^-,\\
\exc(\kk)&=&\min\{\exc^-(\kk),\exc^+(\kk)\},\\
\exc_\ess(\kk)&=&\min\{\exc_\ess^-(\kk),\exc_\ess^+(\kk)\}
.\end{eqnarray}

\subsection{Quasiparticle  systems with the fermionic  superselection rule}

Consider a quasiparticle  system $(H_\fr,P_\fr)$ on the Fock space (\ref{fock}).
Define the {\em fermionic number operator} as
\[ N_\a=\sum_{i\in\cQ_\a}b_i^*(\kk)b_i(\kk).\]
Clearly, the {\em fermionic parity}  $(-1)^{N_\a}$ provides a natural superselection rule.
If $\cH=\cH^+\oplus\cH^-$ denotes the corresponding direct sum decomposition, then the  Hamiltonian and momentum  decompose as
 \begin{equation}
(H_\fr,P_\fr)=(H_\fr^+,P_\fr^+)\oplus(H_\fr^-,P_\fr^-).\label{qrq}\end{equation}
(\ref{qrq}) will be called a {\em two-sector quasiparticle  system}.

Clearly, if we know the dispersion relations
 $I_i\ni\kk\mapsto \omega_i(\kk)$, $i\in\cQ$, then we can determine 
the even and odd energy momentum spectrum of $(H_\fr^+,P_\fr^+)$:
\begin{eqnarray*}
\sp
(H_\fr^+,P_\fr^+)&=&\{(0,\0)\}\\
&&\!\!\cup\ \bigl\{\bigl(\omega_{i_1}(\kk_1)+\cdots+\omega_{i_n}(\kk_n),\kk_1+\cdots+\kk_n\bigr)\ :\\
&&\hskip 5ex
  \hbox{even number of fermions}, n=1,2,3\dots\}^\cl,\\
\sp
(H_\fr^-,P_\fr^-)&=& \bigl\{\bigl(\omega_{i_1}(\kk_1)+\cdots+\omega_{i_n}(\kk_n),\kk_1+\cdots+\kk_n\bigr)\ :\\
&&\hskip 5ex
  \hbox{odd number of fermions}, n=1,2,3\dots\}^\cl
.\end{eqnarray*}

\subsection{Properties of the excitation spectrum of two-sector quasiparticle systems}

Let $(H,P)=(H^+,P^+)\oplus( H^-,P^-)$ be a  two-sector quasiparticle system.
Clearly, we have
\begin{eqnarray}
(0,\0)&\in&\sp(H^+,P^+)\label{pas1a0-}
\end{eqnarray}
because of the Fock vacuum. Here are the  properties of the even and odd excitation spectrum:
\begin{eqnarray}
\sp(H^+,P^+)&=&\sp(H^+,P^+)+\sp(H^+,P^+)\label{pas1a0}
\\
&\supset&
\sp(H^-,P^-)+\sp(H^-,P^-),\label{pas20}\\
\sp(H^-,P^-)&=&\sp(H^-,P^-)+\sp(H^+,P^+).\label{pas30}
\end{eqnarray}

Assume now that
the Hamiltonian is bounded from below.
 Then the Fock vacuum is a translation invariant ground state satisfying $E=0$, so that  the excitation spectrum coincides with the energy-momentum spectrum.
Thus we can rewrite (\ref{pas1a0-})-(\ref{pas30}) as
\begin{eqnarray}
(0,\0)&\in&\sp(H^+-E,P^+)\label{pas1a00},\\
\sp(H^+-E,P^+)&=&\sp(H^+-E,P^+)+\sp(H^+-E,P^+)\label{pas1a00.}
\\
&\supset&
\sp(H^--E,P^-)+\sp(H^--E,P^-),\label{pas200}\\
\sp(H^--E,P^-)&=&\sp(H^--E,P^-)+\sp(H^+-E,P^+).\label{pas300}
\end{eqnarray}

Given (\ref{pas1a00}), (\ref{pas1a00.})-(\ref{pas300}) are equivalent to
\begin{eqnarray}
\Exc^+&\supset&\bigl(\Exc^++\Exc^+\bigr)\cup\bigl(\Exc^-+\Exc^-\bigr),
\label{pas1a00a}
\\
\Exc^-&\supset&\Exc^-+\Exc^+.\label{pas300.}
\end{eqnarray}

If in addition the number of particle species is finite, then
\begin{eqnarray}
\Exc_\ess^+&=&\bigl(\Exc^++\Exc^+\bigr)^\cl\cup \bigl(\Exc^-+\Exc^-\bigr)^\cl,
\label{pas1a000}
\\
\Exc_\ess^-&=&\bigl(\Exc^-+\Exc^+\bigr)^\cl.\label{pas3000}
\end{eqnarray}

\subsection{Two-sector quasiparticle-like spectrum}

Consider now an arbitrary translation invariant  system with two superselection sectors
$(H,P)=(H^+,P^+)\oplus( H^-,P^-)$. We will assume that $H$ is bounded from below and the ground state with energy $E$ is translation invariant and belongs to the sector $\cH^+$.

We will say that the excitation spectrum of $(H^+,P^+)\oplus( H^-,P^-)$ {\em is two-sector quasiparticle-like} if it coincides with the excitation spectrum of a two-sector quasiparticle  system. Such an excitation spectrum has special properties. In particular, it satisfies (\ref{pas1a00})-(\ref{pas300}).

There exists a heuristic 
general argument why realistic translation invariant
quantum systems in thermodynamic limit should satisfy
 (\ref{pas1a00})-(\ref{pas300}).
 It is an obvious modification of the argument given in Subsect.
\ref{Quasiparticle-like excitation spectrum}. 

 Indeed, we need to notice what follows.
 $(-1)^{N_\a}$ is always a superselection rule for  realistic quantum system.
In particular, if we assume that the ground state is nondegenerate, it has to be either bosonic or fermionic. We make an assumption that it is bosonic.

The eigenvectors $\Phi_1$ and $\Phi_2$, discussed in
Subsect.
\ref{Quasiparticle-like excitation spectrum},
 can be chosen to be purely bosonic or fermionic. Using the fact that the ground state is purely bosonic, we see that we can chose the operators $A_1$ and $A_2$ 
to be purely bosonic or fermionic. (That means, they either commute or anticommute with $(-1)^{N_\a}$). Consequently, we have the following possibilities:
\begin{itemize}
\item Both $\Phi_1$ and $\Phi_2$ are bosonic. Then $\Phi_{12}$ is bosonic.
\item Both $\Phi_1$ and $\Phi_2$ are fermionic. Then $\Phi_{12}$ is bosonic.
\item One of $\Phi_1$ and $\Phi_2$  is bosonic, the other is fermionic. Then $\Phi_{12}$ is fermionic. \end{itemize}
This implies (\ref{pas1a00a}) and
(\ref{pas300.}).

\subsection{Bottom of a two-sector quasiparticle-like excitation spectrum}

Suppose again that 
$(H,P)=(H^+,P^+)\oplus(H^-,P^-)$ is a translation invariant 
 system with two superselection sectors.
 We assume that we know  its excitation spectrum.
We would like to describe some criteria to verify whether
it is two-sector quasiparticle-like. These criteria will involve the properties of the bottom of the even and odd excitation spectrum.

\begin{thm} Suppose that the excitation spectrum of $(H^+,P^+)\oplus( H^-,P^-)$ is two-sector quasiparticle-like.
\begin{enumerate}
\item We have the following subadditivity properties:
\begin{eqnarray*}
\exc^-(\kk_1+\kk_2)\leq
\exc^-(\kk_1)+\exc^+(\kk_2),\\
\exc^+(\kk_1+\kk_2)\leq
\exc^-(\kk_1)+\exc^-(\kk_2),\\
\exc^+(\kk_1+\kk_2)\leq
\exc^+(\kk_1)+\exc^+(\kk_2).
\end{eqnarray*}
\item If the number of species of quasiparticles is finite, then we can reconstruct 
$\exc_\ess^-$ and $\exc_\ess^+$ from $\exc^-$ and
$\exc^+$:
\begin{eqnarray*}
\exc_\ess^-(\kk)&=&\inf \{\exc^-(\kk_1)+
\exc^+(\kk_2)\ :\ \kk=\kk_1+\kk_2\},\\
\exc_\ess^+(\kk)&=&\inf \{\exc^+(\kk_1)+
\exc^+(\kk_2),\ \ \ \exc^-(\kk_1)+
\exc^-(\kk_2)\ :\ \kk=\kk_1+\kk_2\}.
\end{eqnarray*}\end{enumerate}
\end{thm}

\subsection{Non-interacting Fermi gas}
\label{Non-interacting Fermi gas}
Let us give a brief discussion of the free
 Fermi gas
 with chemical potential $\mu$ in $d$
dimensions. For simplicity, we will
assume that particles have no internal degrees of freedom such as spin.

The
  Hilbert space of $n$ fermions equals $\Gamma_\a^n\left(L^2(\rr^d)\right)$ 
(antisymmetric square integrable
functions on $(\rr^d)^n$).
Let $\Delta_{(i)}$ denote  the Laplacian $\Delta$ acting on
the $i$th variable.
Then the    Hamiltonian  equals
\begin{equation}
H^{n}=
\sum_{ i=1}^n(-\Delta_{(i)}-\mu).
\label{sch4}\end{equation}
It commutes with the  momentum operator 
\[P^{n}:=\sum_{i=1}^n-\i\nabla_{(i)}.\]

It is convenient to put together various $n$-particle sectors in a single Fock
space 
 \[\Gamma_\a(L^2(\Lambda)):=\mathop{\oplus}\limits_{n=0}^\infty 
\Gamma_\a^n\left(L^2(\Lambda)\right).\]
Then the basic observables are the
 Hamiltonian, the total momentum  and the number operator:
\begin{eqnarray}\nonumber
H&=&\loplus_{n=0}^\infty H^n=\int a_\x^*(-\Delta-\mu)a_\x\d\x,\\
 P&=&\loplus_{n=0}^\infty P^n=-\i\int a_\x^*\nabla_\x a_\x\d\x
,\label{postu1}\\
 N&=&\loplus_{n=0}^\infty n=\int a_\x^* a_\x\d\x
,\nonumber
\end{eqnarray}
where
 $a_\x^*$/ $a_\x$ are the usual fermionic creation/annihilation operators.

The three operators in (\ref{postu1}) describe only a finite number of particles in an
infinite space. We would like to
investigate  homogeneous Fermi gas at a positive density in the thermodynamic limit.
Following the accepted, although somewhat unphysical tradition, 
we first consider our system on
$\Lambda=[-L/2,L/2]^d$, the $d$-dimensional
 cubic box of side length $L$, with periodic boundary conditions.
 Note that the spectrum of the momentum becomes
$ \frac{2\pi }{L}\zz^d$. At the end we let
 $L\to\infty$. 

It is convenient to pass to the momentum representation:
\begin{eqnarray}\label{2B}
H^L 
&=&\sum_{\kk}(\kk^2-\mu) a^*_{\kk}
a_{\kk}\nonumber 
\\
P^L&=&\sum_{\kk}\kk a^*_{\kk}a_\kk,\label{2C}\\
N^L&=&\sum_{\kk}a^*_{\kk}a_\kk,\nonumber\end{eqnarray}
where we used (\ref{postu1})
and $a_\x=L^{-d/2}\sum_\kk \e^{\i \kk\x}a_\kk$. We sum over
$\kk\in \frac{2\pi }{L}\zz^d$.

It is natural to change  the representation of canonical anticommutation
relations
and replace the usual fermionic creation/annihilation operators
by new ones, which
 kill the ground state of the Hamiltonian:
\begin{eqnarray*}
b_\kk^*:&=&a_\kk^*,\ b_\kk:=a_\kk,\ \kk^2>\mu,\\
b_\kk^*:&=&a_\kk,\ b_\kk:=a_\kk^*,\ \kk^2\leq\mu.
\end{eqnarray*}
Then,
\begin{eqnarray*}
 H^L&=&\sum_\kk |\kk^2-\mu|b_\kk^*b_\kk+E^L,\\
 P^L&=&\sum_\kk \kk b_\kk^*
b_\kk ,\\
 N^L&=&\sum_\kk \sgn(\kk^2-\mu) b_\kk^*
b_\kk +C^L
,\label{ham2}\end{eqnarray*}
where 
\begin{eqnarray*}
E^L&=&\sum_{\kk^2\leq\mu} (\kk^2-\mu),\\
C^L&=&\sum_{\kk^2\leq\mu} 1.
\end{eqnarray*}
It is customary to drop the constants $E^L$ and $C^L$.

Set (temporarily) $\omega(\kk)=|\kk^2-\mu|$.
In the case of an infinite space, the above analysis suggests that it is natural to postulate
\begin{eqnarray}
 H&=&\int\omega(\kk)b_\kk^*b_\kk\d\kk,\label{ham3}\\
 P&=&\int \kk b_\kk^*
b_\kk \d\kk,\label{ham3a}\\
 N&=&\int \sgn(\kk^2-\mu) b_\kk^*
b_\kk \d\kk,
\label{ham4}\end{eqnarray}
 as the Hamiltonian, total momentum and number operator of the free Fermi gas from the beginning, instead of (\ref{postu1}).

The operators $b_\kk^*/b_\kk$ can be called quasiparticle creation/annihilation operators and the function $\kk\mapsto\omega(\kk)$ the quasiparticle dispersion relation.  Thus  a quasiparticle is a true particle above the Fermi level and a  hole below the Fermi level.

In Sect. \ref{s5} we describe a version of the BCS theory based on the
 Hartree-Fock-Bogoliubov approximation. This approximation 
 suggests that 
the interacting Fermi gas can be described, at least approximately,  by a Hamiltonian of the form (\ref{ham3}) with a
 dispersion relation $\kk\mapsto\omega(\kk)$ that resembles $|\kk^2-\mu|$, except that
its minimum is strictly positive.

\subsection{Examples of the energy-momentum spectrum}

The energy-momentum spectrum of a Fermi gas described by 
(\ref{ham3}) and (\ref{ham3a}) with various dispersion relations $\omega$ can sometimes  have a curious shape. In the remaining part of
this section we will illustrate this with several examples. We will present
diagrams representing  the energy-momentum spectrum. In the  full
and the odd cases, the dispersion relation
$\omega$ is a singular part of the spectrum and it 
will be denoted by a solid line. In the even case, the dispersion relation
will be denoted by a dotted line.
 We will always consider the spherically symmetric case.

First consider the non-interacting
 Fermi gas, which, as we argued above, has the
dispersion relation $\omega(\kk)=|\kk^2-\mu|$. In dimension $1$ its
energy-momentum spectrum looks quite interesting:

\begin{figure}[!h]
\centering
\includegraphics{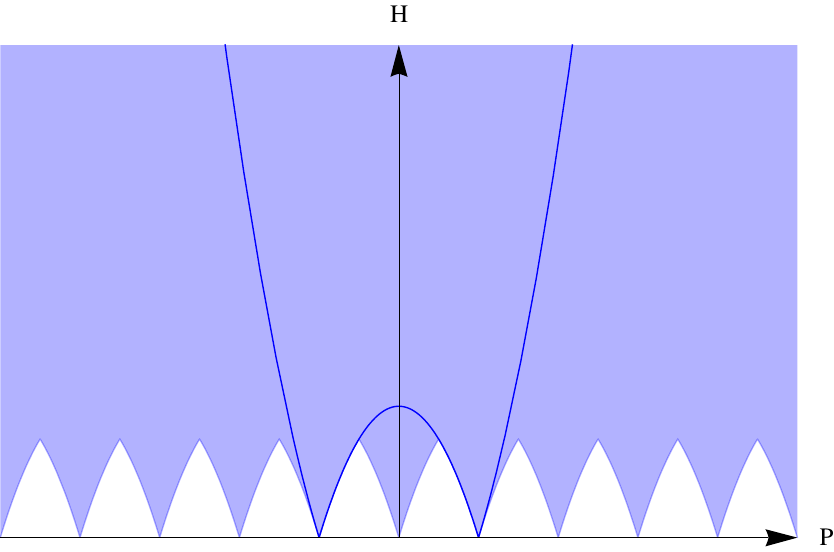}
\label{fig:1}
\caption{$\sp(H,P)$ in the non-interacting case, $d=1$.}
\end{figure}

\begin{figure}[ht]
\centering
\includegraphics{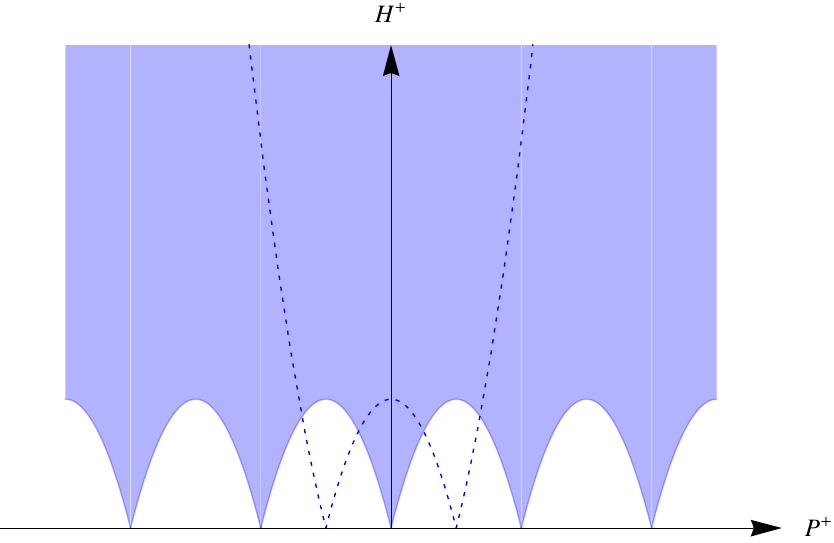}
\label{fig:2}
\caption{$\sp(H^+,P^+)$ in the non-interacting case, $d=1$.}
\end{figure}

\begin{figure}[ht]
\centering
\includegraphics{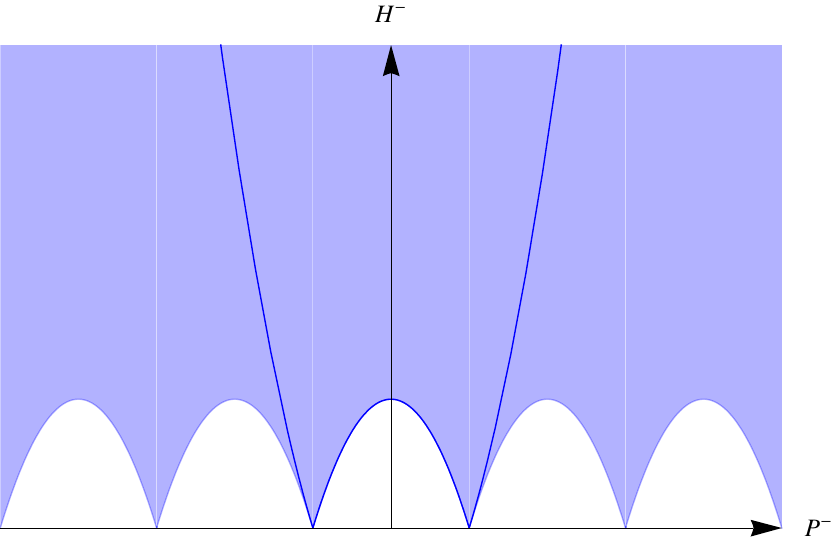}
\label{fig:3}
\caption{$\sp(H^-,P^-)$ in the non-interacting case, $d=1$.}
\end{figure}
\newpage
Clearly, for $d\geq2$ the energy-momentum spectrum is rather boring:
\begin{figure}[!h]
\centering
\includegraphics{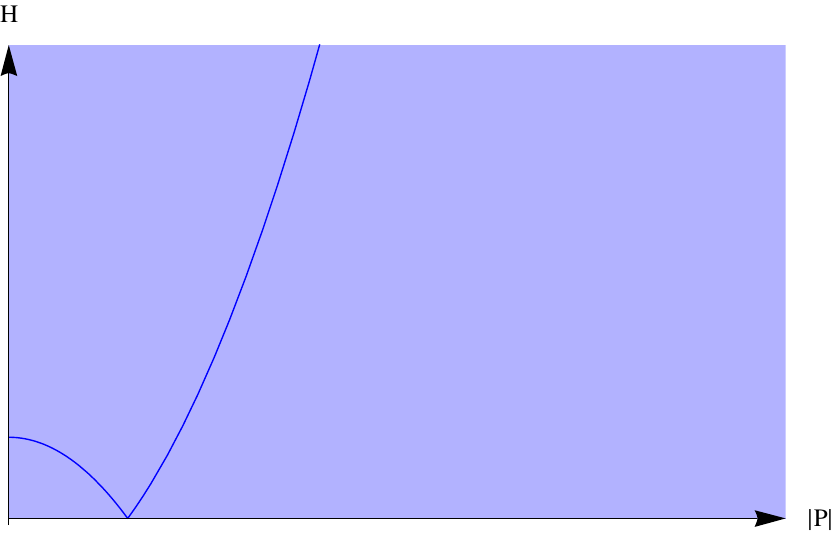}
\label{fig:4}
\caption{$\sp(H,P)$, $\sp(H^+,P^+)$, $\sp(H^-,P^-)$ in the non-interacting case, $d\geq2$.}
\end{figure}

\newpage
In the case of an interacting Fermi gas, we assume that
\begin{eqnarray}
\omega=\sqrt{(\kk^2-\mu)^2+\gamma^2}.\label{dispersion_oddz}
\end{eqnarray}
Calculations presented in Sect. \ref{s5}, in particular equation (\ref{oddz}), suggest that the dispersion relation obtained by the HFB method is qualitatively similar to (\ref{dispersion_oddz}).

\vspace{2cm}


\begin{figure}[!h]
\centering
\includegraphics{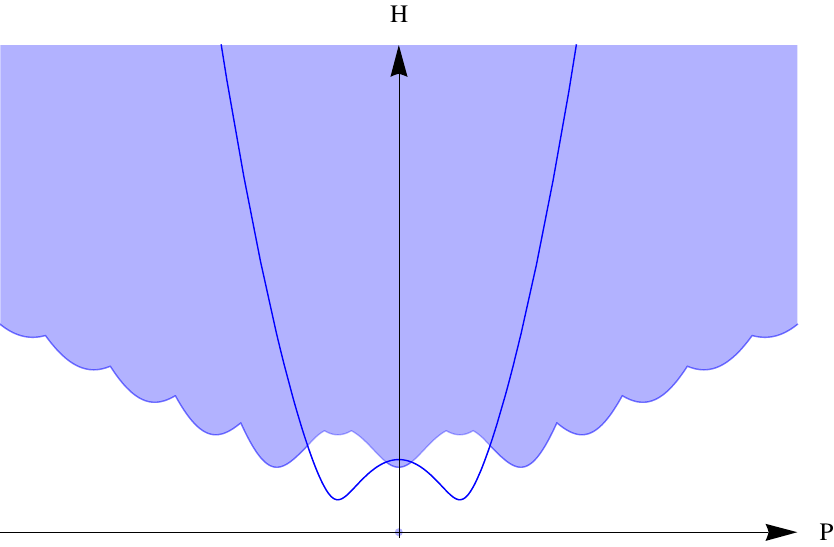}
\label{fig:5}
\caption{$\sp(H,P)$ in the interacting case,  $d=1$.}
\end{figure}

\begin{figure}[!h]
\centering
\includegraphics{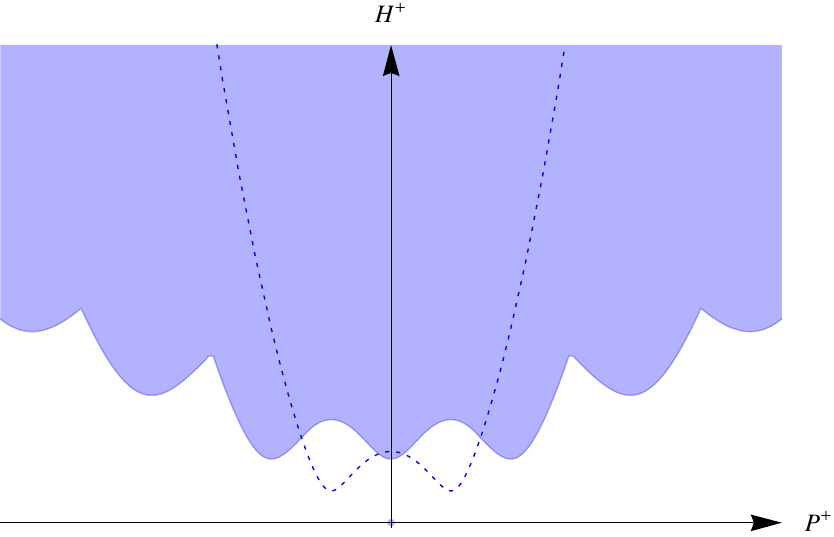}
\label{fig:6}
\caption{$\sp(H^+,P^+)$ in the interacting case, $d=1$.}
\end{figure}

\begin{figure}[!h]
\centering
\includegraphics{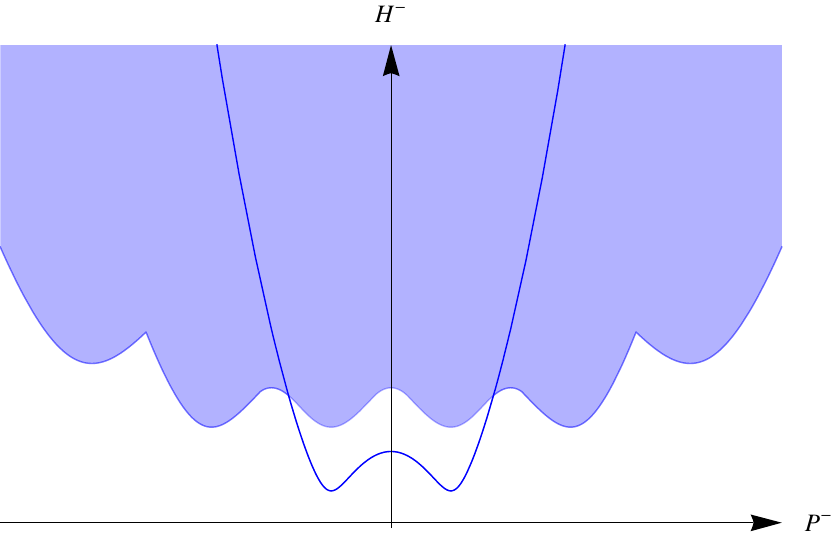}
\label{fig:7}
\caption{$\sp(H^-,P^-)$ in the interacting case, $d=1$.}
\end{figure}

\newpage
\begin{figure}[!h]
\centering
\includegraphics{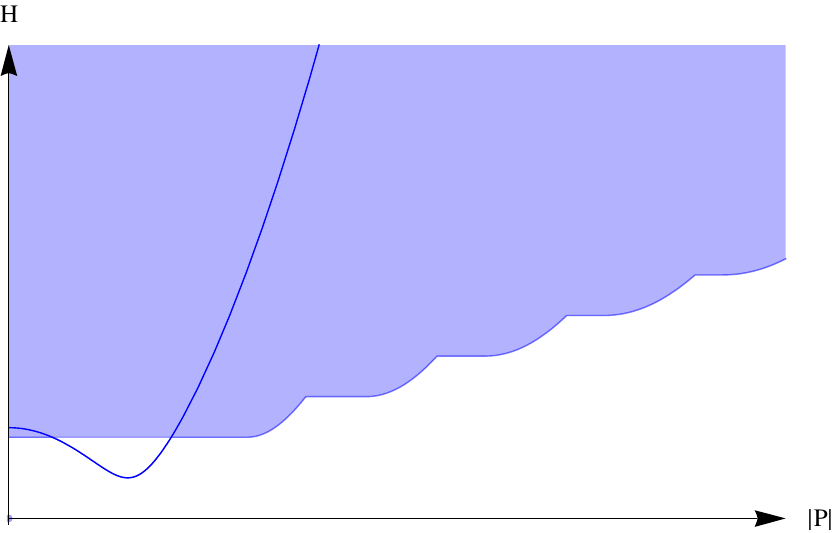}
\label{fig:8}
\caption{$\sp(H,P)$ in the interacting case, $d\geq 2$.}
\end{figure}

\begin{figure}[!h]
\centering
\includegraphics{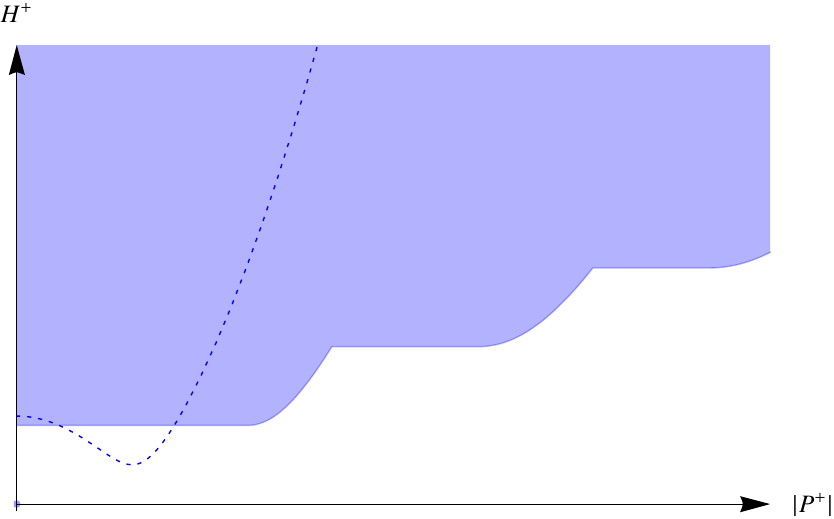}
\label{fig:9}
\caption{$\sp(H^+,P^+)$ in the interacting case, $d\geq 2$.}
\end{figure}

\begin{figure}[!h]
\centering
\includegraphics{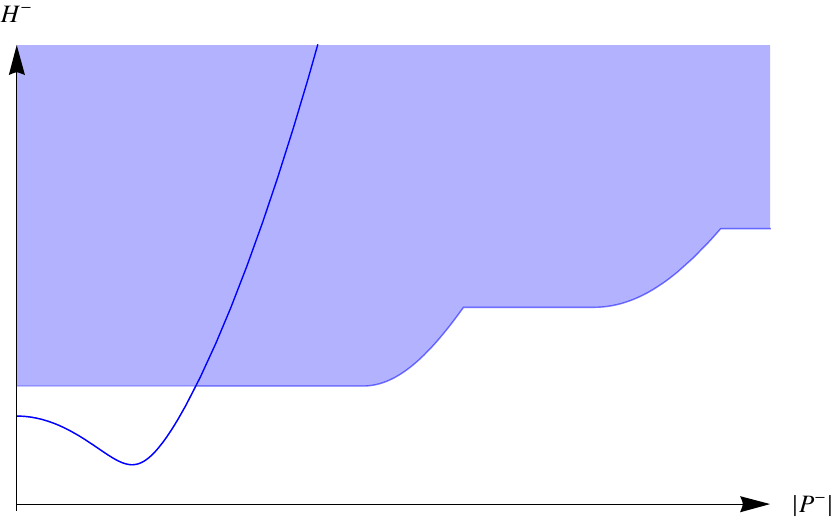}
\label{fig:10}
\caption{$\sp(H^-,P^-)$ in the interacting case, $d\geq 2$.}
\end{figure}

\newpage

Again, the case $d=1$ differs from
$d\geq2$.  However, in all dimensions
the energy gap and the critical velocity are
 strictly positive.

\section{The model and conjectures}
In this section we describe some  classes of interacting models
that  seem to be relevant for condensed matter physics. We also
formulate conjectures about these models that we think
are suggested by ``physical intuition''.

Internal degrees of freedom of particles, such as spin, play an important  role in
fermionic systems. They are in particular crucial
 in the BCS approach. Therefore, we will take them
into account. We will assume that they are described
by a finite dimensional Hilbert space $\cc^m$. Thus the one-particle space of
the system is $L^2(\rr^d,\cc^m)$.

\subsection{1-particle energy}

The kinetic energy of one particle
including its chemical potential is given by 
a self-adjoint 
 operator
 $T$ on $L^2(\rr^d,\cc^m)$.
We use the following notation for its integral kernel: for $\Phi\in
L^2(\rr^d,\cc^m)$, 
\[(T\Phi)_{i_1}(\x_1)=
\sum_{i_2}\int T_{i_1,i_2}(\x_1,\x_2)
\Phi_{i_2}(\x_2)\d \x_2.\]
 We assume that $T$ is a self-adjoint and
translation invariant  one-body operator. Clearly,
\begin{eqnarray*}
T_{i_1,i_2}(\x_1,\x_2)&=&\bar{ T_{i_2,i_1}(\x_2,\x_1)}\\
&=&T_{i_1,i_2}(\x_1+\y,\x_2+\y).\end{eqnarray*}
The first identity expresses the hermiticity and the second the
translation invariance of $T$.

We will sometimes assume that $T$ is {\em real}, that is,
invariant with respect to
the complex conjugation. This means
that $T_{i_1,i_2}(\x_1,\x_2)$ are real.
An example of a real 1-particle energy is
\[T_{ij}=\bigl(-\frac{1}{2m_i}\Delta-\mu_i\bigr)\delta_{i,j},\] where the $i$th
``spin'' has the mass $m_i$  and the chemical
potential $\mu_i$.

If the operator $T$ has the form 
\[T_{i,j}(\x_i,\x_j)=t(\x_i,\x_j)\delta_{i,j},\]
for some function $t$ satisfying
\begin{eqnarray*}
t(\x_1,\x_2)&=&\bar{t(\x_2,\x_1)}\\
&=&t(\x_1+\y,\x_2+\y),\end{eqnarray*}
then we will say that $T$ is {\em spin-independent}.

Clearly, the 1-particle energy can be written as
\[T_{i,j}(\x_1,\x_2)=(2\pi)^{-d}\int\tau_{i,j}(\kk)\e^{\i\kk(\x_1-\x_2)}\d\kk.\]

If it is real, then $\tau_{i,j}(\kk)=\tau_{i,j}(-\kk)$

If it is spin independent, then
\[\tau_{i,j}(\kk)=\tau(\kk)\delta_{i,j}.\]

In the real spin-independent case we have $\tau(\kk)=\tau(-\kk)$.
\subsection{Interaction}

 The
interaction of the Fermi gas 
will be described by a
2-body operator $V$. It acts on the antisymmetric 2-particle
space as
\[(V\Phi)_{i_1,i_2}(\x_1,\x_2)=
\sum_{i_3,i_4}\int\int
V_{i_1,i_2,i_3,i_4}(\x_1,\x_2,\x_3,\x_4)\Phi_{i_4,i_3}(\x_4,\x_3)\d \x_3\d \x_4,\]
where  $\Phi\in \Gamma_\a^2\left( L^2(\rr^d,\cc^m)\right)$.
We will assume that it is
 self-adjoint translation invariant.
Its  integral kernel satisfies 
\begin{eqnarray*}
V_{i_1,i_2,i_3,i_4}(\x_1,\x_2,\x_3,\x_4)&=&
-V_{i_2,i_1,i_3,i_4}(\x_2,\x_1,\x_3,\x_4)
\\
&=&-V_{i_1,i_2,i_4,i_3}(\x_1,\x_2,\x_4,\x_3)\\
&=&\overline{V_{i_4,i_3,i_2,i_1}(\x_4,\x_3,\x_2,\x_1)}\\
&=&V_{i_1,i_2,i_3,i_4}(\x_1+\y,\x_2+\y,\x_3+\y,\x_4+\y).
\end{eqnarray*}
The first two identities express the antisymmetry of the interaction, the
third --  its hermiticity and the fourth -- its translation invariance.
 We also  assume 
that $V(\x_1,\x_2,\x_3,\x_4)$ decays for large differences of its
 arguments sufficiently fast.

We will sometimes assume that $V$  are {\em real}, that means, they are invariant with respect to
the complex conjugation. This means
 $V_{i_1,i_2,i_3,i_4}(\x_1,\x_2,\x_3,\x_4)$ is real.

 We will say 
that the operator $V$ is {\em spin independent}
if there exists a function
 $v(\x_1,\x_2,\x_3,\x_4)$
such that \begin{eqnarray*}
&&V_{i_1,i_2,i_3,i_4}(\x_1,\x_2,\x_3,\x_4)\\&=&
\frac12\bigl(v(\x_1,\x_2,\x_3,\x_4)
\delta_{i_1,i_4}\delta_{i_2,i_3}
-v(\x_1,\x_2,\x_4,\x_3)\delta_{i_1,i_3}\delta_{i_2,i_4}\bigr),
\end{eqnarray*}
Note that
\begin{eqnarray*}
v(\x_1,\x_2,\x_3,\x_4)&=&
v(\x_2,\x_1,\x_4,\x_3)\\
&=&\overline{v(\x_4,\x_3,\x_2,\x_1)}\\
&=&v(\x_1+\y,\x_2+\y,\x_3+\y,\x_4+\y).
\end{eqnarray*}

It will be convenient to write the Fourier transform of $V$ as follows
\begin{eqnarray*}
&&V(\x_1,\x_2,\x_3,\x_4)\\&=&
(2\pi)^{-4d}
\int\e^{\i \kk_1\x_1+\i\kk_2\x_2-\i \kk_3\x_3-\i\kk_4\x_4}
 Q(\kk_1,\kk_2,\kk_3,\kk_4)\\&&\times\delta
(\kk_1+\kk_2-\kk_3-\kk_4)\d\kk_1\d\kk_2\d\kk_3\d\kk_4,\end{eqnarray*}
where
$ Q(\kk_1,\kk_2,\kk_3,\kk_4)$ is a function defined on the subspace
$\kk_1+\kk_2=\kk_3+\kk_4$. (Thus we could drop, say, $\kk_4$ from its
arguments; we do not do it for the sake of the symmetry of formulas).
Clearly,
\begin{eqnarray*}
Q_{i_1,i_2,i_3,i_4}(\kk_1,\kk_2,\kk_3,\kk_4)&=&
-Q_{i_2,i_1,i_3,i_4}(\kk_2,\kk_1,\kk_3,\kk_4)
\\
&=&-Q_{i_1,i_2,i_4,i_3}(\kk_1,\kk_2,\kk_4,\kk_3)\\
&=&\overline{Q_{i_4,i_3,i_2,i_1}(\kk_4,\kk_3,\kk_2,\kk_1)}.
\end{eqnarray*}

If we assume that the interaction is real, then
\begin{eqnarray*}
 Q_{i_1,i_2,i_3,i_k}(\kk_1,\kk_2,\kk_3,\kk_4)&=&\bar{ Q_{i_1,i_2,i_3,i_k}
(-\kk_1,-\kk_2,-\kk_3,-\kk_4)}.
\end{eqnarray*}

If  we assume that the interaction is spin-independent, then 
\begin{eqnarray*} 
Q_{i_{1}i_{2}i_{3}i_{4}}(\kk_1,\kk_2,\kk_3,\kk_4) & = &\frac12\bigl(
q(\kk_1,\kk_2,\kk_3,\kk_4)\delta_{i_{1}i_{4}}\delta_{i_{2}i_{3}}
-q(\kk_1,\kk_2,\kk_4,\kk_3)\delta_{i_{1}i_{3}}\delta_{i_{2}i_{4}}\bigr),
\end{eqnarray*}
for some function $q$ defined on $\kk_1+\kk_2=\kk_3+\kk_4$
satisfying
\begin{eqnarray*}
q(\kk_1,\kk_2,\kk_3,\kk_4)&=&
q(\kk_2,\kk_1,\kk_4,\kk_3)\\
&=&\overline{q(\kk_4,\kk_3,\kk_2,\kk_1)}.
\end{eqnarray*}

In the real spin-independent case we have in addition
\[q(\kk_1,\kk_2,\kk_3,\kk_4)=
\bar{q(-\kk_1,-\kk_2,-\kk_3,-\kk_4)}.\]

For example,
 a 2-body potential $V(\x)$ such that $V(\x)=V(-\x)$ corresponds to the real spin-independent interaction with
\begin{eqnarray*}
v(\x_1,\x_2,\x_3,\x_4)
&=& V(\x_1-\x_2)
\delta(\x_1-\x_4)\delta(\x_2-\x_3),\\
q(\kk_1,\kk_2,\kk_3,\kk_4)&=&\int\d\q
\hat V(\q)\delta(\kk_1-\kk_4-\q)\delta(\kk_2-\kk_3+\q).
\end{eqnarray*}

\subsection{$n$-body Hamiltonian}

The
$n$-body  Hamiltonian of the homogeneous Fermi gas 
acts on  the Hilbert space $\Gamma_\a^n\left(L^2(\rr^d,\cc^m)\right)$ 
(antisymmetric square integrable
functions on $(\rr^d)^n$ with values in $(\cc^m)^{\otimes n}$).
Let $T_{(i)}$ denote  the operator $T$ acting on
the $i$th variable and 
$V_{(ij)}$ denote the  operator $V$ acting on the $(ij)$th pair of
variables. 
The full $n$-body Hamiltonian  equals
\begin{equation}
H^{n}=
\sum_{1\leq i\leq n}T_{(i)}+
\sum_{1\leq i<j\leq n}V_{(ij)}.
\end{equation}
It commutes with the  momentum operator 
\[P^{n}:=\sum_{i=1}^n-\i\nabla_{\x_i}.\]

\subsection{Putting system in a box}

As discussed already in the previous section, to
investigate homogeneous Fermi gas at positive density in thermodynamic limit
it is convenient to put the system
on a box
$\Lambda=[-L/2,L/2]^d$
with periodic boundary conditions. This means in particular that
the kinetic energy is replaced by
\[T^L(x_1,x_2)=\frac{1}{L^d}
\sum_{\kk \in \frac{2\pi}{L}\mathbb{Z}^d}\e^{\i \kk\cdot( \x_1-x_2)}\tau(\kk)
,\]
and
the  potential $V$ is replaced  by
\begin{eqnarray*}\label{interakt}
&&V^L(\x_1,\x_2,\x_3,\x_4)=\\
&=&\frac{1}{L^{3d}}\sum_{\kk_1,\dots,\kk_4
\in \frac{2\pi}{L}\mathbb{Z}^d,\ \ \kk_1+\kk_2=\kk_3+\kk_4}
\e^{\i\kk_1 \cdot \x_1+\i\kk_2\x_2-\i\kk_3\x_3-
\i\kk_4\x_4} Q(\kk_1,\kk_2,\kk_3,\kk_4).
\end{eqnarray*}
Note that $V^L$ is periodic with respect to the domain
$\Lambda$, and $V^L(\x)\to V(\x)$ as $L\to\infty$. 
 The system  on a torus  is described by the Hamiltonian
\begin{equation}
H^{L,n}=
\sum_{1\leq i\leq n}T_{(i)}^L+
\sum_{1\leq i<j\leq n}V_{(ij)}^L
\label{sch}\end{equation}
acting on the space $\Gamma_\a^n\left(L^2(\Lambda,\cc^m)\right)$.

\subsection{Grand-canonical Hamiltonian of the Fermi gas}
\label{s2b}

It is convenient to put all the $n$-particle spaces into a single Fock space
 \[\Gamma_\a(L^2(\Lambda,\cc^m)):=\mathop{\oplus}\limits_{n=0}^\infty 
\Gamma_\a^n\left(L^2(\Lambda,\cc^m)\right)\]
with the
Hamiltonian 
\begin{eqnarray*}
H^L&:=&\mathop{\oplus}\limits_{n=0}^\infty H^{L,n}\\
&=&\int a_{\x,i_1}^* T_{i_1,i_2}^L(\x_{i_1}-\x_{i_2}) a_{\x,i_2}\d
\x_1\d\x_2 \nonumber\\
&&+\frac12\int\int a_{\x_1,i_1}^*a_{\x_2,i_2}^*
V_{i_1,i_2,i_3,i_4}^L(\x_1,\x_2,\x_3,\x_4) a_{\x_3,i_3} a_{\x_4,i_4}\d
\x_1\d\x_2\d\x_3\d\x_4,
\end{eqnarray*}
where $a_{\x,i}$, $a_{\x,\i}^*$ are the usual fermionic annihillation and
creation operators.
The second quantized
momentum and number operators are defined as
\begin{eqnarray*}
P^L&:=&
\mathop{\oplus}\limits_{n=0}^\infty P^{n,L}=-\i\int
a_{\x,i}^*\nabla_\x^La_{\x,i} \d\x,\\
N^L&:
=&\mathop{\oplus}\limits_{n=0}^\infty n=\int a_{\x,i}^*a_{\x,i}\d\x
.\end{eqnarray*}
Above we use the summation convention. In what follows we will usually omit the
indices.

In the momentum representation,
\begin{eqnarray}
H^L 
&=&\sum_{\kk}\tau(\kk)a^*_{\kk}
a_{\kk}\nonumber \\
&&+\frac{1}{2L^d}\sum_{\kk_1+\kk_2=\kk_3+\kk_4}Q(\kk_1,\kk_2,\kk_3,\kk_4)
a^*_{\kk_1}a^*_{\kk_2}a_{\kk_3}a_{\kk_4},\label{resa5}
\\
P^L&=&\sum_{\kk}\kk a^*_{\kk}a_\kk,\nonumber
\\
N^L&=&\sum_{\kk}a^*_{\kk}a_\kk.\nonumber
\end{eqnarray}

In the spin-independent case, the interaction equals
\[\frac{1}{2L^d}\sum_{\kk_1+\kk_2=\kk_3+\kk_4}q(\kk_1,\kk_2,\kk_3,\kk_4)
a^*_{\kk_1,i}a^*_{\kk_2,j}a_{\kk_3,j}a_{\kk_4,i}
\]
In the case of a (local) potential, it is
\[\frac{1}{2L^d}\sum_{\kk,\kk',\q}\hat V(\q)
a^*_{\kk+\q,i}a^*_{\kk'-\q,j}a_{\kk',j}a_{\kk,i}.
\]

$H^{L,\pm}(\kk)$ will denote the operator $H^L$ restricted to the
subspace $P^L=\kk$, $(-1)^{N^L}=\pm1$.

\subsection{Infimum of the excitation spectrum}
\label{Infimum of the excitation spectrum}
For a large class of potentials the finite volume Hamiltonians $H^L$ are bounded from below and have a discrete spectrum. 

The ground state energy  is defined as
\begin{equation}\label{lagrtra}
E^{L}={\rm inf\; sp}H^{L}.
\end{equation} 
 For $\kk\in  \frac{2\pi }{L}\zz^d$ 
 we define the {\em infimum of the excitation spectrum in the even/odd
   sector in finite volume}: 
\begin{eqnarray*}\label{defexcit}
\epsilon^{L,+}(\kk)&:=&
\inf\sp H^{L,+}(\kk){-}E^{L},\ \kk\neq0,\\
\epsilon^{L,+}(\0)&:=&
\inf\big(\sp (H^{L,+}(\0){-}E^{L})\backslash\{0\}\big),\\
\epsilon^{L,-}(\kk)&:=&
\inf\sp H^{L,-}(\kk){-}E^{L}.
\end{eqnarray*}


For $\kk\in\rr^d$, we would like 
 to define the {\em  infimum of the excitation spectrum in 
thermodynamic
  limit}. To this end, first we define its finite volume version in a
``window'' given by $\delta>0$:
\begin{eqnarray*}
\epsilon^{L,\delta,\pm}(\kk)
&:=&\inf
\{\epsilon^{L,\pm}(\kk'_L)\ :\ 
\kk'_L\in      \frac{2\pi}{L}\zz^d,\ |\kk-\kk'_L|<\delta\}.
\end{eqnarray*}
Then we set
\begin{eqnarray*}
\epsilon^\pm(\kk)&=&
\sup_{\delta >0}\left(\liminf_{L\to\infty}\left(\epsilon^{L,\delta,\pm}(\kk)
\right)\right).
\end{eqnarray*}



Let us now formulate our conjectures about $\epsilon^\pm$.

\begin{conjecture}\label{conj1a}
We expect that for a large class of  potentials with attractive interactions
the following statements hold true: 
\begin{enumerate} 
\item The functions
 $\rr^d\ni \kk\mapsto \epsilon^{\pm}(\kk)
\in \rr$  are continuous.
\item Let $\kk\in{\mathbb R}^d$.
 Let $(\kk_s,L_s)
\in\frac{2\pi}{L_s}\zz^d\times [0,\infty)$
obey  $\kk_s\to \kk$,
$L_s\to\infty$. Then
 $\epsilon^{L_s,\pm}(\kk_s) \to \epsilon^{\pm}(\kk)$.
\item If $d\geq2$, then
\begin{eqnarray*}
\inf_\kk\min\bigl(\epsilon^{-}(\kk),\epsilon^+(\kk)\bigr)&=&:\varepsilon>0.
\end{eqnarray*}
\item If $d\geq2$, then
\begin{eqnarray*}
\inf_{\kk\neq\0}\frac{\min\bigl(\epsilon^{-}(\kk),\epsilon^+(\kk)\bigr)}{|\kk|}&=&:c_\crit>0.
\end{eqnarray*}
\item We have the following subadditivity properties:
\begin{eqnarray*}
\epsilon^-(\kk_1+\kk_2)\leq
\epsilon^-(\kk_1)+\epsilon^+(\kk_2),\\
\epsilon^+(\kk_1+\kk_2)\leq
\epsilon^-(\kk_1)+\epsilon^-(\kk_2),\\
\epsilon^+(\kk_1+\kk_2)\leq
\epsilon^+(\kk_1)+\epsilon^+(\kk_2).
\end{eqnarray*}
\end{enumerate}\label{conje}
\end{conjecture}

To motivate the above conjecture, consider
a
model Hamiltonian
\begin{equation}H= \sum_{i\in\cQ}\int_{I_i}\omega_i(\kk)b^*_{\kk,i}
b_{\kk,i}\d\kk,\label{exci}\end{equation}
where $b_{\kk,i}$,  $b_{\kk,i}^*$ are (fermionic, but possibly also bosonic) annihillation/creation
operators and $I_i\ni \kk\mapsto \omega_i(\kk)$ are continuous functions defined on closed subsets $I_i\subset \rr^d$.  For $\kk\in\rr^d$, let
$\omega_{\min}(\kk)$ be the lowest dispersion relation defined as in (\ref{mino}).
Assume that
\begin{eqnarray*}\varepsilon&:=&\inf_\kk\omega_{\min}(\kk)>0,\\
c_\crit&:=&\inf_{\kk\neq\0}\frac{\omega_{\min}(\kk)}{|\kk|}>0,
\end{eqnarray*}
which is suggested by the HFB approximation, see Sect. \ref{s5}.
Then the  infimum of the even/odd 
 excitation spectrum of the Hamiltonian $H$ equals
$\epsilon^\pm(\kk)=\sah_{\omega_{\min}}^\pm(\kk)$ and has the properties described in
Conjecture \ref{conje}.

Note that in this conjecture we expect statements (3) and (4) to be true only in $d\geq 2$. This is due to an argument based on the Galilean covariance in a box with periodic boundary conditions in one dimension explained in Sect. II B of \cite{CDZ}. It is valid both for bosons and fermions.

\subsection{Isolated quasiparticle shells}
\label{Excited quasiparticle shells}

 The quadratic part of the model Hamiltonian (\ref{non-})
obtained as the result of the HFB approximation
 involves 
 $m$ fermionic quasiparticles (corresponding to the dimension of the ``internal subspace'' $\cc^m$). Its excitation spectrum
will
 contain ``lacunas'' above its infimum separated by at most $m$ shells.
 In
this subsection we try to formulate an additional conjecture
that takes these lacunas into account. This is more difficult
than the conjectures from the previous subsection. It is also more dubious.

 For $\kk\in  \frac{2\pi }{L}\zz^d$ and $j,n\in\nn$,
 we define the {\em$j$th   shell in finite volume in the $n$-body case} 
\begin{eqnarray*}\nu_j^{L,n,+}(\kk)&:=&
\hbox{the $j$th lowest eigenvalue 
of }H^{L,n,+}(\kk){-}E^{L},\ \ \kk\neq\0,\\
\nu_j^{L,n,+}(\0)&:=&
\hbox{the $j+1$st lowest eigenvalue 
of }H^{L,n,+}(\0){-}E^{L},\\
\nu_j^{L,n,-}(\kk)&:=&
\hbox{the $j$th lowest eigenvalue 
of }H^{L,n,-}(\kk){-}E^{L}.\end{eqnarray*}
(Of course, when 
counting eigenvalues  we take into account their multiplicity).


Let $\nn^+:=\{0,2,4,\dots\}$ and $\nn^-:=\{1,3,5,\dots\}$.
For $\kk\in\rr^d$, we would like 
 to define the {the $j$th  shell in 
thermodynamic
  limit}.  To this end, first we define its finite volume version in a
``window'' given by $\delta>0$:
\begin{eqnarray*}
&&\nu_j^{L,\delta,+}
(\kk)\\
&:=&\inf_{\kk'_L\in
      \frac{2\pi}{L}\zz^d,\; |\kk-\kk'_L|<\delta,\; n\in \nn^\pm}
\{\nu_j^{L,n}(\kk'_L),\ \inf\sp H^{L,n}-E^L<\delta\}.\\
\end{eqnarray*}
Then we set
\begin{eqnarray*}
\nu_j^\pm(\kk)&=&
\sup_{\delta >0}\left(\liminf_{L\to\infty}\left(\nu_j^{L,\delta,\pm}(\kk)
\right)\right ).
\end{eqnarray*}

Clearly,
\begin{eqnarray*}
\nu_1^\pm(\kk)&=&\epsilon^\pm(\kk),\\
\nu_j^{\pm}(\kk)&\leq&\nu_{j+1}^{\pm}(\kk).\end{eqnarray*}


Set
\[\epsilon_\ess^\pm(\kk):=\sup\{\nu_j^\pm(\kk)\ :\ j=1,2,\dots\}.\]
Let us now formulate the conjectures about $\epsilon_\ess^\pm$.

\begin{conjecture}\label{conj1a=}
We expect that for a large class of attractive potentials
the following statements hold true: 
\begin{enumerate} 
\item The functions
 $\rr^d\ni \kk\mapsto \nu_j^\pm(\kk), \ \epsilon_\ess^\pm(\kk)
\in \rr_+$
 are continuous.
\item
 Let $(\kk_s,L_s,n_s)
\in\frac{2\pi}{L_s}\zz^d\times [0,\infty[\times\nn^\pm$
obey  $\kk_s\to \kk$, $\inf\sp H^{L_s,n_s}-E^{L_s}\to0$,
$L_s\to\infty$. Then
 \[\nu_j^{L_s,n_s,\pm}(\kk_s)\to \nu_j^{\pm}(\kk).\]
A similar property holds for $\epsilon_\ess^\pm$.
\item $\epsilon_\ess^\pm$ are related to $\epsilon^\pm$ as follows:
\begin{eqnarray*}
\epsilon_\ess^-(\kk)&=&\inf \{\epsilon^-(\kk_1)+
\epsilon^+(\kk_2)\ :\ \kk=\kk_1+\kk_2\},\\
\epsilon_\ess^+(\kk)&=&\inf \{\epsilon^+(\kk_1)+
\epsilon^+(\kk_2),\ \ \ \epsilon^-(\kk_1)+
\epsilon^-(\kk_2)\ :\ \kk=\kk_1+\kk_2\}.
\end{eqnarray*}
\end{enumerate}
\end{conjecture}

To justify this conjecture, 
let us note first that it is consistent with the spectral properties of
the
model Hamiltonian
(\ref{exci}) if we assume that the number of quasiparticles is finite.

We can try to be more precise.
We expect that the functions $\nu_j$ stabilize. In other words, for a certain $m^\pm$ and $j\geq m^\pm$,
all $\nu_j^\pm$ are equal to one another, and hence equal to $\epsilon_\ess^\pm$.
Then it is natural to guess that the functions $\omega_j$ that appear in the model Hamiltonian (\ref{exci}) and correspond to bosonic, resp. fermionic quasiparticles coincide with $\nu_j^\pm$ for $j\leq m^\pm$.

Note that the HFB approximation, described in the next section,  suggests that  $m^+=0$ and  $m^-=m$, where $m$ is the number of internal degrees of freedom. In particular, this would  mean that all quasiparticles are fermionic. 
This conjecture is probably too strong. One cannot exclude that the interaction  leads to a formation of   quasiparticles consisting of an even number of fermions. Such quasiparticles would be of course bosonic.

\section{The 
Hartree-Fock-Bogoliubov approximation applied to homogeneous 
  Fermi gas}  
\label{s5}

One can try to
compute the excitation spectrum of the Fermi gas
by approximate methods. Historically, the
first computation of this sort is due to Bardeen-Cooper-Schrieffer. 
 In its original
version, the BCS method involved a replacement of quadratic fermionic
operators with 
bosonic ones. We will use the approach based on a Bogoliubov
rotation of fermionic variables, which is commonly called the
Hartree-Fock-Bogoliubov method. Its main idea is to 
minimize the
energy in the so-called fermionic Gaussian
 states -- states obtained by a Bogoliubov
rotation from the fermionic Fock vacuum. The minimizing state will define new
creation/annihilation operators. We express the Hamiltonian in the new 
creation/annihilation operators and drop all higher order terms. This defines
a new Hamiltonian, that we expect to give an approximate description of low
energy part of the excitation spectrum.

\subsection{The rotated Hamiltonian}
One can start the HFB method with a rotation of
the fermionic creation/annihilation operators. 
For any $\kk$ this corresponds to a substitution
  \begin{eqnarray}
&a_\kk^*=c_\kk b_\kk^*+s_\kk b_{-\kk},&
a_\kk=\bar c_\kk b_\kk+\bar s_\kk b_{-\kk}^*,\label{rota}\end{eqnarray}
where $c_\kk$ and  $s_\kk$ are matrices on $\cc^m$ satisfying
\begin{eqnarray}\label{susti1}
c_\kk c_\kk^*+s_\kk s_\kk^*&=&1,\\ \label{susti2}
c_\kk s_{-\kk}^\t+s_\kk c_{-\kk}^\t&=&0.
\end{eqnarray}
($*$ denotes the hermitian conjugation,
$\#$ denotes the transposition and $\overline{\cdot}$ denotes the complex conjugation).

Here (\ref{susti1}) guarantees that $[a_\kk^*,a_\kk]_+=1$,
(\ref{susti2}) guarantees that $[a_\kk^*,a_{-\kk}^*]_+=0$. Note that 
 $[a_\kk^*,a_{\kk'}]_+=0$,  $[a_\kk^*,a_{-\kk'}^*]_+=0$, $\kk\neq\kk'$
 are satisfied automatically.



For a sequence
$\frac{2\pi}{L}{\mathbb Z}^d\ni \kk\mapsto \theta_\kk$ 
with values in matrices on $\cc^m$ such that
 $\theta_\kk=\theta_{-\kk}$, set
\begin{equation}
U_{\theta}:=\prod_\kk\e^{-\frac12
\theta_\kk a_\kk^*a_{-\kk}^*+\frac12
    \theta_\kk^* a_\kk a_{-\kk}}.\label{uthe}\end{equation}
It is well known that for an appropriate sequence $\theta$
we have
\[U_\theta^*
a_\kk
U_\theta=b_\kk,\ \
\ \ 
U_\theta^*
a_\kk^*U_\theta
=b_\kk^*.\]
Note also that $U_{\theta}$
is the general form of an even
 Bogoliubov transformation commuting with $P^L$.

In this section  we drop the superscript $L$, writing eg. $H$ for $H^L$.
The Hamiltonian (\ref{resa5}) after the substitution (\ref{rota}) and
 the Wick ordering equals
\begin{align}\nonumber
   H 
&=
B
\\&+\frac12\sum_\kk O(\kk)b_\kk^*b_{-\kk}^*+
\frac12\sum_\kk\bar O(\kk)b_{-\kk} b_{\kk} +\sum_\kk
D(\kk)b_\kk^*b_\kk\nonumber
\\&
+\hbox{terms  higher order in {\it b}'s}\label{non}.
\end{align}
  Here are explicit formulas for $B$,
 $D(\kk)$ and $O(\kk)$:
\begin{eqnarray}
B&=&
\sum_\kk\tau(\kk) s_\kk \bar s_\kk
\nonumber\\&&
+\frac{1}{2L^d}\sum_{\kk,\kk'}
 Q(\kk,-\kk,-\kk',\kk') s_{\kk}c_{-\kk}\bar c_{-\kk'}\bar s_{\kk'}\nonumber\\
&&+\frac{1}{L^d}
\sum_{\kk,\kk'}Q(\kk,\kk',\kk',\kk)s_{\kk}s_{\kk'}
\bar s_{\kk'}\bar s_{\kk}
;\nonumber
\end{eqnarray}
\begin{eqnarray}
O(\kk)&=&
2\tau(\kk) c_\kk\bar s_\kk\nonumber\\
&&+\frac{1}{L^d}\sum_{\kk'}
Q(\kk',-\kk',-\kk,\kk)s_{\kk'}c_{-\kk'}\bar s_{-\kk}\bar
s_{\kk}\nonumber\\ 
 &&+\frac{1}{L^d}\sum_{\kk'}
Q(\kk,-\kk,-\kk',\kk')c_{\kk}c_{-\kk}\bar c_{-\kk'}\bar s_{\kk'}
\nonumber\\
&&+\frac{4}{L^d}\sum_{\kk'} Q(\kk,\kk',\kk',\kk)c_{\kk}s_{\kk'}
\bar s_{\kk'}\bar
s_{\kk}\nonumber\\ 
D(\kk)&=&\tau(\kk)
c_\kk\bar c_\kk-\bigl(\tau(\kk) s_{-\kk} \bar s_{-\kk}\bigr)^{\rm T}\nonumber\\
&&+\frac{1}{L^d}\sum_{\kk'}
 Q(\kk',-\kk',-\kk,\kk)s_{\kk'}c_{-\kk'}\bar s_{-\kk}\bar
c_{\kk}\nonumber\\ 
 &&+\frac{1}{L^d}\sum_{\kk'}
 Q(\kk,-\kk,-\kk',\kk')c_{\kk}s_{-\kk}\bar c_{-\kk'}\bar s_{\kk'}
\nonumber\\
&&+\frac{2}{L^d}\sum_{
\kk'}
 Q(\kk,\kk',\kk',\kk)
c_{\kk}s_{\kk'}\bar s_{\kk'}\bar
c_{\kk}
\nonumber\\
&&-\frac{2}{L^d}\sum_{
\kk'}\bigl(
 Q(-\kk,\kk',\kk',-\kk)
 s_{-\kk}s_{\kk'}\bar s_{\kk'}
\bar s_{-\kk}\bigr)^{\rm T}.
\nonumber
\end{eqnarray}

Note that the       formulas for $B$, $O(\kk)$ and $D(\kk)$ are
written in a special notation, whose aim is to avoid putting a big
number of internal indices.
 The matrices $c_\kk$ and $s_\kk$ have two internal indices: right and left. 
 We sum over the right internal indices, whenever we sum over the
 corresponding momenta. The left internal indices are contracted with the
 corresponding indices of $\tau$ or $ Q$.
The superscript $\rm T$ stands for the transposition (swapping
the indices).

\subsection{Minimization over Gaussian states}
\label{Minimization over Gaussian states}
Let $\Omega$ denote the  vacuum vector. $\Omega_{\theta}:=U_{\theta}^*\Omega$
is the general form of an even fermionic Gaussian  vector of zero momentum.
Clearly,
\begin{eqnarray}
(\Omega_{\theta}|H \Omega_{\theta})&=&B,\label{min1}
\\ (b_\kk^*\Omega_{\theta}|H 
b_\kk^*\Omega_{\theta})
&=&B+D(\kk).\label{min2}\end{eqnarray}
Therefore, we obtain rigorous bounds
\begin{eqnarray*}
E&\leq& B,\\
E+\epsilon^{-}(\kk)&\leq &B+\inf D(\kk).
\end{eqnarray*}

We would like to find a fermionic Gaussian  vector that minimizes $B$ -- the expectation value of
$H$.
We assume that there exists a stationary point $(\tilde{s}_\kk,\tilde{c}_\kk)$
of $B$ considered as a function of $c$ and $s$. 
 Bogoliubov transformations form a group,
hence the neighbourhood of the stationary point can be expressed in the
following way: 
\begin{equation}
\left[ \begin{array}{cc}
  c_\kk & s_\kk \\
  \bar s_\kk & \bar c_\kk\\       
     \end{array}  \right]=
\left[ \begin{array}{cc}
  \tilde{c}_\kk & \tilde{s}_\kk \\
  \bar{\tilde s}_\kk & \bar{\tilde c}_\kk\\       
     \end{array}  \right]
\left[ \begin{array}{cc}
  c_\kk' & s'_\kk \\
  \bar s'_\kk & \bar c_\kk'\\       
     \end{array}  \right].
\end{equation}
This means (including internal indices) that
\begin{eqnarray*}
c_{il, \kk}&=&\tilde{c}_{im, \kk}c'_{ml, \kk}+\tilde{s}_{im, \kk}\bar{s'}_{ml,
  \kk}, \nonumber \\
\bar{c}_{il, \kk}&=&\bar{\tilde{s}}_{im, \kk}s'_{ml, \kk}+\bar{\tilde{c}}_{im,
  \kk}\bar c'_{ml, \kk}, \nonumber \\
s_{il, \kk}&=&\tilde{c}_{im, \kk}s'_{ml, \kk}+\tilde{s}_{im, \kk}\bar c'_{ml,
  \kk},  \nonumber \\
\bar{s}_{il, \kk}&=&\bar{\tilde{s}}_{im, \kk}c'_{ml, \kk}+\bar{\tilde{c}}_{im, \kk}\bar{s'}_{ml, \kk}. \nonumber
\end{eqnarray*}
We enter the above formulas into the expressions for $B, O(\kk)$ and
$D(\kk)$. 

We can always multiply $c_\kk$ and $s_\kk$ by a unitary matrix without changing
the Gaussian state. Hence, we can assume that
\begin{eqnarray}
c_\kk'&=&\sqrt{1-(s_\kk')^*s_\kk'}.
\label{pas1}\end{eqnarray}
Since $s'$ is a complex function we can treat $s'$ and $\bar{s}'$ as
independent variables.
$c_\kk=\tilde c_\kk$, $s_\kk= \tilde s_\kk$ corresponds to
$s'=0$, $\bar s'=0.$
Because of (\ref{pas1}), we have
\begin{eqnarray*}
\frac{\partial}{\partial s'_{\kk}}c_\kk'\Big|_{\substack{s'=0
    \\ \bar{s'}=0}}&=&0,\\ 
\frac{\partial}{\partial \bar s'_{\kk}} c_\kk'\Big|_{\substack{s'=0
    \\ \bar{s'}=0}}&=&0.
\end{eqnarray*}
Then, for
example, taking the first term of $B$ one gets 
\begin{align*}
\frac{\partial}{\partial s'_{rt, \kk'}}\sum_{\kk}\tau_{\alpha\beta,
  \kk}s_{\alpha\alpha', \kk}\bar{s}_{\beta \alpha',
  \kk}\Big|_{\substack{s'=0 \\ \bar{s'}=0}}
=\tau_{\alpha\beta, \kk'}\tilde{c}_{\alpha r,
  \kk'}\bar{\tilde{s}}_{\beta t, \kk'} ,
\end{align*} 
which equals the first term of $O(\kk)$ at $c=\tilde{c}$ and
$s=\tilde{s}$. Calculating other terms of $B$ one finally gets 
\begin{equation}
\frac{\partial B}{\partial s'}\Big|_{\substack{s'=0
    \\ \bar{s'}=0}}=\frac12O(\kk)|_{\substack{c=\tilde{c}
    \\ s=\tilde{s}}}. \label{zerowanie} 
\end{equation}
Thus the minimizing procedure is equivalent to $O(\kk)=0$.
This result is a special case of a more general fact discussed in \cite{DNS} where it is called the Beliaev Theorem \cite{Bel}.

Thus, if we choose the Bogoliubov transformation according to the minimization procedure, the Hamiltonian equals
\begin{align}
   H 
&=
B+\sum_\kk
D(\kk)b_\kk^*b_\kk
+\hbox{terms  higher order in {\it b}'s}\label{non-}.
\end{align}

In the case of the model interaction considered by Bardeen-Cooper-Schrieffer, described in many texts, eg. in \cite{FW},
the minimization of $B$ yields  a dispersion relations that has a positive energy gap and a positive critical velocity uniformly as $L\to\infty$, that is,
\begin{eqnarray}\inf_\kk D(\kk)>0,
&&\inf_{\kk\neq\0}\frac{D(\kk)}{|\kk|}>0.\label{resa4}
\end{eqnarray}
 This phenomenon is probably much more general.  In particular, we expect that it is true for a large class of real, spin-independent and attractive interactions. In what follows we provide computations that seem to support this claim.

Note that the reality and spin-independence of the interactions leads to a considerable computational simplification. By an attractive interaction we mean an interaction, which in some sense, described later on, is negative definite.

Let us assume in addition that higher order terms in (\ref{non-}) are in some sense negligible. Then formally $H$ is approximated by a quadratic Hamiltonian $B+\sum_\kk
D(\kk)b_\kk^*b_\kk$ whose dispersion relation  has a strictly positive energy gap and critical velocity.
We view this as an argument in favor of Conjectures \ref{conj1a} and
\ref{conj1a=}.

\subsection{Reality condition}

Let us first apply the assumption about the reality of the interaction. In this
case, it is natural to assume that the trial vector is real as well. This
means that we impose the conditions
\[\bar c_\kk=c_{-\kk}, \ \ \ \bar s_\kk =s_{-\kk}.\] This allows us to
simplify the formulas for $B$, $D(\kk)$ and $O(\kk)$:
\begin{eqnarray}\nonumber
B&=&\sum_\kk\tau(\kk) s_\kk\bar s_\kk\nonumber\\
&&+\frac{1}{2L^d}\sum_{\kk,\kk'}
 Q(\kk,-\kk,-\kk',\kk') s_\kk\bar c_\kk
c_{\kk'}
\bar s_{\kk'}\nonumber\\ 
&&+\frac{1}{L^d}\sum_{\kk,\kk'}Q(\kk,\kk',\kk',\kk)\bar s_{\kk}\bar s_{\kk'}
s_{\kk'}
s_{\kk},\nonumber\\[3ex]
O(\kk)&=&
2\tau(\kk) c_\kk\bar s_\kk\nonumber\\
&&+\frac{1}{L^d}\sum_{\kk'}
 Q(\kk,-\kk,-\kk',\kk')(c_{\kk}\bar c_{\kk}-s_\kk\bar s_\kk)
c_{\kk'}\bar
s_{\kk'}\nonumber\\ 
&&+\frac{4}{L^d}\sum_{\kk'}Q(\kk,\kk',\kk',\kk)c_{\kk}s_{\kk'}
\bar s_{\kk'}\bar
s_{\kk},\nonumber\\[3ex]
D(\kk)&=&\tau(\kk)
(c_\kk\bar c_\kk- s_{\kk} \bar s_{\kk})\nonumber\\
&&+\frac{2}{L^d}\sum_{\kk'}
 Q(\kk,-\kk,-\kk',\kk')c_{\kk}\bar s_{\kk}c_{\kk'}\bar
s_{\kk'}\nonumber\\ 
&&+\frac{2}{L^d}\sum_{
\kk'}
 Q(\kk,\kk',\kk',\kk)
(c_{\kk}s_{\kk'}\bar s_{\kk'}\bar
c_{\kk}
-s_\kk s_{\kk'}\bar s_{\kk'}\bar s_\kk).
\nonumber
\end{eqnarray}

\subsection{Spin $\frac12$ case}
Assume that the ``spin space'' is $\cc^2$ and the Hamiltonian is spin
independent. 
We make the BCS ansatz:
\begin{eqnarray*}
c_\kk&=&\cos\theta_\kk\left[\begin{array}{cc}
{1}&{0}\\{0}&{1}\end{array}\right],\\
s_\kk&=&\sin\theta_\kk\left[\begin{array}{cc}
{0}&{1}\\{-1}&{0}\end{array}\right],
\end{eqnarray*}
where, keping in mind the reality condition, the parameters $\theta_{\kk}$ are
real.
Then 
\begin{eqnarray*}
B&=&\sum_\kk\tau(\kk)(1-\cos2\theta_\kk)\\
&&+\frac{1}{4L^d}\sum_{\kk,\kk'}\alpha(\kk,\kk')\sin2\theta_\kk\sin2\theta_{\kk'}\\
&&+\frac{1}{4L^d}\sum_{\kk,\kk'}\beta(\kk,\kk')
(1-\cos2\theta_\kk)(1-\cos2\theta_{\kk'}),\end{eqnarray*}
where
\begin{eqnarray*}
\alpha(\kk,\kk')&:=&\frac12\bigl(
q(\kk,-\kk,-\kk',\kk')+q(-\kk,\kk,-\kk',\kk')\bigr),\\
\beta(\kk,\kk')&=&2q(\kk,\kk',\kk',\kk)
-q(\kk',\kk,\kk',\kk).
\end{eqnarray*}
Note that
\[\alpha(\kk,\kk')=\alpha(\kk',\kk),\ \ \ 
\beta(\kk,\kk')=\beta(\kk',\kk).\]
In particular, in the case of local potentials we have
\begin{eqnarray*}
\alpha(\kk,\kk')&:=&\frac12
\bigl(\hat V(\kk- \kk')+\hat V(\kk+ \kk')\bigr),
\\
\beta(\kk,\kk')&=&2\hat V(\0)-\hat V
(\kk-\kk').
\end{eqnarray*}
We further compute:
\begin{eqnarray*}
O(\kk)&= &\bigl(\delta(\kk)\cos2\theta_{k}+
\xi(\kk)\sin2\theta_{\kk}\bigr)\left[\begin{array}{cc}
{0}&{1}\\{-1}&{0}\end{array}\right],\\
D(\kk)& = & (\xi(\kk)\cos2\theta_k-\delta(\kk)\sin2\theta_{k}) \left[\begin{array}{cc}
{1}&{0}\\{0}&{1}\end{array}\right],
\end{eqnarray*} 
where
\begin{eqnarray*}
\delta(\kk)&=& \frac{1}{2L^d}\sum_{\kk'}\alpha(\kk,\kk')\sin2\theta_{\kk'},\\ 
\xi(\kk) & = &
\tau(\kk)+\frac{1}{2L^d}\sum_{\kk'}\beta(\kk,\kk')(1-\cos2\theta_{\kk'}).
\end{eqnarray*}

We are looking for a minimum of $B$. To this end, we first analyze critical points of $B$. We compute the derivative of $B$:
\[\partial_{2\theta_\kk}B=\delta(\kk)\cos2\theta_\kk+\xi(\kk)\sin2\theta_\kk.\]
The condition $\partial_{2\theta_\kk}B=0$, or equivalently $O(\kk)=0$,
has many solutions. We can have
\begin{eqnarray} \sin2\theta_\kk=0,&&\cos2\theta_\kk=\pm1,\label{pqi1}
\\
\hbox{or}&&\nonumber\\
\sin2\theta_\kk=-\epsilon_\kk\frac{\delta(\kk)}{\sqrt{\delta^2(\kk)+
\xi^2(\kk)}}\neq0,&&
\cos2\theta_\kk=\epsilon_\kk\frac{\xi(\kk)}{\sqrt{\delta^2(\kk)+
\xi^2(\kk)}},
\end{eqnarray}
where $\epsilon_\kk=\pm1$. 

In particular, there are many solutions with  all $\theta_\kk$ satisfying (\ref{pqi1}). They correspond to Slater determinants and have a fixed number of particles. The solution of this kind that minimizes $B$ is called the {\em normal} or {\em Hartree-Fock solution}.

One expects that under some conditions the normal solution is not the global minimum of $B$. More precisely, one expects that a global minimum is reached by a configuration satisfying
\begin{eqnarray}
\sin2\theta_\kk=-\frac{\delta(\kk)}{\sqrt{\delta^2(\kk)+
\xi^2(\kk)}},&&
\cos2\theta_\kk=\frac{\xi(\kk)}{\sqrt{\delta^2(\kk)+
\xi^2(\kk)}},\label{resa1}
\end{eqnarray}
where at least some of $\sin 2\theta_\kk$ are different from $0$. It is sometimes called a {\em superconducting solution}. 
In such a case we  get
\begin{eqnarray}
D(\kk)=\sqrt{\xi^2(\kk)+\delta^2(\kk)}\left[\begin{array}{cc}
{1}&{0}\\{0}&{1}\end{array}\right]. \label{oddz}
\end{eqnarray}
Thus we obtain a positive dispersion relation. One can expect that it is strictly positive, since otherwise the two functions $\delta$ and $\xi$ would have  a coinciding zero, which seems unlikely. 
Thus we expect that the dispersion relation $D(\kk)$ has a positive energy gap.

If the interaction is small, then $\xi(\kk)$ is close to $\tau(\kk)$ and $\delta(\kk)$  is small. This implies that $D(\kk)$ is close to $|\tau(\kk)|$.
If $\tau(\kk)$ has a critical velocity for large $\kk$ and $D(\kk)$ has an energy gap, then this implies that $D(\kk)$ also has a critical velocity.

In other words,  we expect that for a large class of interactions if the minimum of $B$ is reached at a superconducting state, then $D(\kk)$ satisfies (\ref{resa4}).

We will not study conditions guaranteeing that a superconducting solution minimizes the energy in this  paper. Let us only remark that such conditions involve some kind of negative definiteness of the quadratic form  $\alpha$ -- this is what we vaguely indicated by saying that the interaction is attractive. Indeed,
 multiply the definition of $\delta(\kk)$ with $\sin2\theta_\kk$ and sum it up over $\kk$. We then obtain
\begin{equation}
\sum_\kk\sin^22\theta_\kk\sqrt{\delta^2(\kk)+\xi^2(\kk)}
=-\frac{1}{2L^d}\sum_{\kk,\kk'}\sin2\theta_\kk\alpha(\kk,\kk')\sin2\theta_{\kk'}
.\label{resa9}\end{equation}
The left hand side of (\ref{resa9}) is positive.
This means that the quadratic form given by the kernel $\alpha(\kk,\kk')$ has to be negative at least at the vector given by $\sin2\theta_\kk$.

Let us also indicate why one expects that the solution corresponding to  (\ref{resa1}) is a  minimum of $B$.
We compute the second derivative:
\begin{eqnarray}\nonumber
\partial_{2\theta_\kk}\partial_{2\theta_{\kk'}}B&=&
\delta_{\kk,\kk'}\bigl(-\sin2\theta_\kk\delta(\kk)+\cos2\theta_{\kk}\xi(\kk)
\bigr)
\\\nonumber
&&+\frac{1}{2L^d}\alpha(\kk,\kk')\cos2\theta_\kk\cos2\theta_{\kk'}\\\label{resa2}
&&+\frac{1}{2L^d}\beta(\kk,\kk')\sin2\theta_\kk\sin2\theta_{\kk'}.
\end{eqnarray}
Substituting (\ref{resa1}) to the first term on the right of (\ref{resa2}) gives
\[\delta_{\kk,\kk'}\sqrt{\delta^2(\kk)+\xi^2(\kk)},\]
which is positive definite. One can hope that  the other two terms in the second derivative of $B$ do not spoil its positive definiteness.


\begin{thebibliography}{}
\bibitem{BCS}
      Bardeen, J., Cooper, L. N., Schrieffer, J. R., \emph{Theory of superconductivity}, Phys. Rev. 108 (1957) 1175
\bibitem{Bel}  Beliaev, S. T.: Effect of pairing correlations on nuclear properties, Mat.-Fys. Skr. Danske
Vid. Selsk 31 (11), 1959
\bibitem{BRT}  Black, C.~T..,  Ralph, D.~C., and  Tinkham, M.:
Spectroscopy of the Superconducting Gap in Individual Nanometer-Scale
Aluminum Particles
Phys. Rev. Lett. 76  (1996) 688
\bibitem{Bog} Bogoliubov, N. N.,  J. Phys. (USSR) {\bf 9}, 23 (1947);
  J. Phys. USSR  {\bf 11}, 23 (1947), reprinted in D. Pines {\em The Many-Body
    Problem } (New York, W.A. Benjamin 1962) 

\bibitem{BBZKT} N.N. Bogolyubov (jr), J.G. Brankov, V.A. Zagrebnov, A.M. Kurbatov,
and IM.S. Tonchev: Some classes of exactly soluble models of
problems in quantum statistical mechanics:
the method of the approximating Hamiltonian,
Russian Math. Surveys 39:6 (1984), 1-50

\bibitem{BR1} Bratteli, O., Robinson D. W.: {\em Operator Algebras and
Quantum Statistical Mechanics, Volume 1}, Springer-Verlag, Berlin, 1987, second 
edition. 
\bibitem{BR2} Bratteli, O., Robinson D. W., 
1996: {\em Operator Algebras and
Quantum Statistical Mechanics, Volume 2}, Springer-Verlag, Berlin, second  edition.


\bibitem{CDZ}  Cornean, H., Derezi\'{n}ski, J.,  Zi\'{n}, P.:
 On the infimum of the energy-momentum spectrum of a homogeneous Bose gas,
 J. Math. Phys. 50,  (2009)   062103

\bibitem{CS}
 Critchley, R. H. and Solomon,  A. I.:
A Variational Approach to
Superfluidity, Journal of Statistical Physics,  14, p.  381-393, 1976
\bibitem{De1}  Derezi\'{n}ski, J.: Asymptotic completeness of long-range $N$-body quantum systems, Ann. of Math. 138, 427-476 (1993)
\bibitem{De2}  Derezi\'{n}ski, J.:
Asymptotic completeness in quantum field theory. A class of Galilei covariant models, Rev. Math. Phys. 10 (1998) 191-233
\bibitem{DG1} Derezi\'{n}ski, J., G\'erard,~C.: Asymptotic completeness in quantum field theory.\\
Massive Pauli-Fierz Hamiltonians, Rev. Math. Phys. 11 (1999) 383-450.
\bibitem{DNS}  Derezi\'{n}ski, J., Napi\'{o}rkowski, M.,  Solovej, J. P.:
 On the minimization of Hamiltonians over pure gaussian states, preprint, arXiv:1102.2931 
\bibitem{FW} Fetter, A. L., Walecka, J. D.: Quantum theory of
  many-particle systems, McGraw-Hill Book Company 1971
\bibitem{FGS} 
Fr\"ohlich,~J., Griesemer,~M., and Schlein,~B.: Asymptotic completeness for Compton scattering.
Comm. Math. Phys. 252  (2004) 415--476
\bibitem{GJ} Glimm, J., Jaffe, A., 1987: {\em Quantum Physics. A Functional 
Integral Point of View}, second edition, Springer-Verlag, New-York.
\bibitem{Jost} Jost,~R: {\em  The general theory of quantized fields},
  AMS, Providence, Rhode Island 1965  
\bibitem{M} Maris, H.~J., "Phonon-phonon interactions in liquid helium",
  Rev. Mod. Phys. {\bf 49}, 341 (1977) 
\bibitem{RS}
      Ring, P., Schuck, P., \emph{The Nuclear Many-body Problem}, Springer-Verlag, New York, 1980
\bibitem{Sch}
 Schwarz,~A.~S: {\em Mathematical foundations of quantum field theory}
 (Russian) Nauka
 1975  
\bibitem{Sei}
Seiringer, R.: The Excitation Spectrum for Weakly Interacting Bosons, Commun. Math. Phys.
306, 565–578 (2011).



\end{thebibliography}
\end{document}